\documentclass[10pt,twocolumn,journal,transmag]{IEEEtran}
\usepackage{xcolor}
\usepackage{subfigure}
\usepackage{algorithm}
\usepackage{algorithmic}
\usepackage{amsmath} 
\usepackage{graphicx}
\usepackage{multicol}
\usepackage{multirow}
\graphicspath{{./images/}}
\usepackage{ulem}

\usepackage{float} 
\usepackage{soul} 
\usepackage{subfigure}
\usepackage{cite}
\usepackage[numbers,sort&compress]{natbib}
\usepackage{hyperref}

\ifCLASSINFOpdf

\else

\fi
\begin{document}

\title{Language-Independent Approach for Automatic Computation of Vowel Articulation Features in Dysarthric Speech Assessment}

\author{\IEEEauthorblockN{Yuanyuan Liu \href{https://orcid.org/0000-0002-1839-4728}{\includegraphics[scale=0.02]{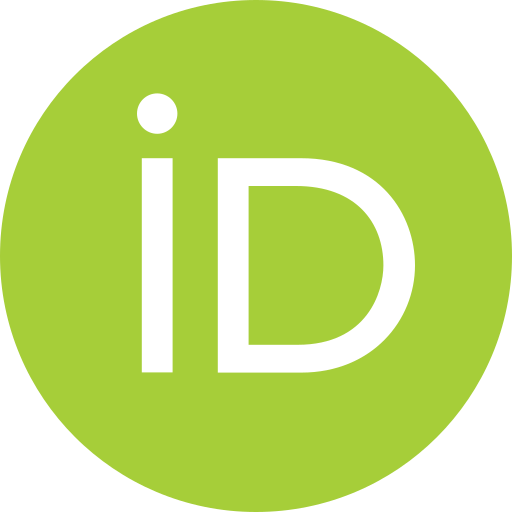}}\IEEEauthorrefmark{1},
 Nelly Penttilä\IEEEauthorrefmark{2}, 
 Tiina Ihalainen\IEEEauthorrefmark{2}, 
 Juulia Lintula\IEEEauthorrefmark{2}, 
 Rachel Convey\IEEEauthorrefmark{2}, 
 Okko Räsänen\IEEEauthorrefmark{1,3}}
\IEEEauthorblockA{\IEEEauthorrefmark{1}Unit of Computing Sciences, Tampere University, Finland}
\IEEEauthorblockA{\IEEEauthorrefmark{2}Faculty of Social Sciences, Tampere University, Finland}
\IEEEauthorblockA{\IEEEauthorrefmark{3}Dept. Signal Processing and Acoustics, Aalto University, Finland}
\thanks{\copyright 2021 IEEE. Personal use of this material is permitted. Permission from IEEE must be obtained for all other uses, in any current or future media, including reprinting/republishing this material for advertising or promotional purposes, creating new collective works, for resale or redistribution to servers or lists, or reuse of any copyrighted component of this work in other works.}}

% The paper headers
\markboth{IEEE/ACM TRANSACTIONS ON AUDIO, SPEECH, AND LANGUAGE PROCESSING, Vol.29, 2021}%
{Shell \MakeLowercase{\textit{et al.}}: Bare Demo of IEEEtran.cls for IEEE Transactions on Magnetics Journals}

\IEEEtitleabstractindextext{%
\begin{abstract}
Imprecise vowel articulation can be observed in people with Parkinson's disease (PD). Acoustic features measuring vowel articulation have been demonstrated to be effective indicators of PD in its assessment. Standard clinical vowel articulation features of vowel working space area (VSA), vowel articulation index (VAI) and formants centralization ratio (FCR), are derived the first two formants of the three corner vowels /a/, /i/ and /u/. Conventionally, manual annotation of the corner vowels from speech data is required before measuring vowel articulation. This process is time-consuming. The present work aims to reduce human effort in clinical analysis of PD speech by proposing an automatic pipeline for vowel articulation assessment. The method is based on automatic corner vowel detection using a language universal phoneme recognizer, followed by statistical analysis of the formant data. The approach removes the restrictions of prior knowledge of speaking content and the language in question. Experimental results on a Finnish PD speech corpus demonstrate the efficacy and reliability of the proposed automatic method in deriving VAI, VSA, FCR and F2i/F2u (the second formant ratio for vowels /i/ and /u/). The automatically computed parameters are shown to be highly correlated with features computed with manual annotations of corner vowels. In addition, automatically and manually computed vowel articulation features have comparable correlations with experts' ratings on speech intelligibility, voice impairment and overall severity of communication disorder. Language-independence of the proposed approach is further validated on a Spanish PD database, PC-GITA, as well as on TORGO corpus of English dysarthric speech.

\end{abstract}

}
\maketitle
\IEEEdisplaynontitleabstractindextext
\IEEEpeerreviewmaketitle

\section{Introduction}
\label{sec:intro}

\IEEEPARstart{P}{arkinson’s} disease (PD), the second most common neurodegenerative disease, has a wide range of symptoms, including characteristic movement disorders and non-motor symptoms on sleep, mental and cognitive performance \cite{PD_neorodisease}. 
Among all mentioned symptoms, speech impairment has been demonstrated to be an early indicator and a valuable marker of disease progression and treatment efficacy of PD \cite{Rusz_2013}. It is said that up to $90\%$ of people with PD develop hypokinetic dysarthria \cite{PD_dysarthria, Ramig_2008}, which is a perceptually distinct motor speech disorder \cite{Duffy_2012}. 
The manifestations of hypokinetic dysarthria can include impaired phonation, imprecise articulation, reduced variability of pitch and loudness, and other prosodic disturbances related to speech rate, stress and pauses \cite{Duffy_2012, Rusz_2013}. Due to these deficits, speech intelligibility of people with PD could be degraded.

One of the frequent signs of PD is the presence of a progressively imprecise articulation \cite{Ramig_2008}. People with PD are likely to fail to reach the articulatory targets or sustain the articulation for a sufficient duration, which is known as articulatory undershoot \cite{Duffy_2012}. The articulatory undershoot together with reduced range and rate of articulatory movement result in inaccurate articulation of vowels and consonants. Previous studies show that articulation disorders can be quantitatively measured with acoustic analysis, which serves as a reliable, objective and non-invasive tool for detection and progression monitoring of PD \cite{HarelBrianT2004AcoP, Rusz_2013}. In this context, vowel articulation in PD speech has attracted researchers' attention \cite{VSA_Liu_2005, SKODDA_2011}, since vowel clarity has been shown to be a powerful indicator of speech intelligibility \cite{AAVS_clarity_2016, vowel_intelligibility_2017}.

Speech and language pathologists are trained to identify and differentiate communication disorders by using auditory perceptual judgement and acoustic analysis \cite{Penttila_2018}. Similarly, speech intelligibility is commonly assessed perceptually, with the help of articulation tests and manual phonetic transcriptions \cite{Kent_2020}. As said before, vowel articulation is a reliable indicator of speech intelligibility, but it requires manual annotations. In fact, according to our knowledge, all the works on studying vowel articulation in dysarthric speech have used vowel segments that have been manually extracted from the speech stimuli \cite{VSA_Liu_2005, VAI, SKODDA_2011, Sapir_2011, Skodda_2012, speech_tasks_2013, Mou_VAI}. The process of manual annotation could be precise, but it is also time-consuming and requires the annotator(s) to have basic understanding of speech analysis, annotation tools (e.g., Praat \cite{praat}) and the language at hand. This is a burden on clinical work, and also limits the scalability of automated patient screening and follow-up using clinically interpretable features. However, recent technical developments in automatic speech recognition (ASR) have been successfully involved in feature extraction in automatic assessment of various types of pathological speech \cite{Zlotnik2015RandomFP, williamson2015segment,LIU_2019}. This raises the question whether analysis of vowel articulation could also be fully automated in order to support clinical practice.

Given this background, the main purpose of the present study is to develop an automatic and language-independent method for vowel articulation measurement in terms of acoustic features.  
Inspired by the recent successes of using ASR for feature extraction in automatic pathological speech assessment  \cite{Zlotnik2015RandomFP, williamson2015segment,LIU_2019}, a universal phone/phoneme recognizer is adopted to detect speech frames representative of corner vowel articulation, followed by statistical analysis of the formant frequencies across the detected frames. 

We demonstrate the reliability and efficacy of the proposed method in automatic detection of corner vowels related speech frames using a Finnish corpus of read speech from people with PD as well as healthy controls. In addition, we demonstrate that the automatically computed vowel articulations are correlated with expert assessment of speech intelligibility, voice impairment and overall severity of communication disorder. We also provide evidence for language-independence of our approach by testing the same system on a Spanish PD corpus as well as an English corpus of dysarthric speech. The results show that the automatically computed vowel articulation parameters have significant differences between speech from control speakers and PD/dysarthric speakers. A number of the automatically computed vowel articulation parameters are also moderately correlated with UPDRS (Unified Parkinson's Disease Rating Scale) and UPDRS-speech scores in the Spanish PD corpus as well as the overall dysarthric severity level for the English dysarthria corpus.

This article is organized as follows: Section \ref{sec:relwork} describes the background literature in more detail. Section \ref{sec:method} introduces the system framework, algorithm for automatic corner-vowel-related frames selection and vowel articulations computation. Section \ref{sec:corpus} describes the speech corpora used in the present work. Experimental results are presented and discussed in Section \ref{sec:result} and \ref{sec:discuss}. This article is concluded in Section \ref{sec:conclude}.

\subsection{Related work}
\label{sec:relwork}
When using acoustic analysis, features such as fundamental frequency, formants, intensity and spectrum are computed from recordings of different speech tasks, such as sustained phonation of vowels, sentence repetition, reading and monologue \cite{speech_tasks_2013, Rusz_2013}. With the acoustic parameters, pattern recognition techniques can be applied to discriminate speech from control and PD groups \cite{Nonotn_2014}. Correlations between the acoustic parameters and severity scores on multiple dimensions like voice, speech and motor disorder have been investigated as well \cite{Rusz_2013}.

In \cite{Nonotn_2014}, $13$ articulatory features were extracted from PD-related dysarthric speech of diadochokinetic `/pa-ta-ka/'-repetition based on an automatic algorithm to detect initial burst, vowel onset and occlusion. These articulatory features describe the voice quality, coordination of laryngeal and supralaryngeal activity, precision of consonant articulation, tongue movement, occlusive weakening and speech timing. The features on consonant articulation were found to be the most sensitive indicators of PD-related dysarthria.

Vowel articulation is also central to PD-related dysarthric speech due to the inherent coupling between PD-associated hypokinesia and the reach and accuracy of articulatory gestures. 
In terms of acoustics, different vowels can be distinguished by their formant frequencies, which are resonances formed by the vocal tract \cite{Sunberg_2017}. Specifically, the first formant (F1) corresponds to the height of tongue body in articulation, whilst the second formant (F2) corresponds to the frontness/backness of tongue body \cite{english_vowels}. 
To qualitatively evaluate the clarity (or, conversely, undershoot) of vowel articulation, formant related features are widely used, such as vowel articulation space (VSA), vowel articulation index (VAI), formant centralization ratio (FCR), the second formant ratio for vowels /i/ and /u/ (F2i/F2u) and so on \cite{VAI, FCR}. These features are computed from formants of corner vowels /a/, /i/ and /u/, which are most commonly used in human languages and represent the extreme positions of the speaker's articulatory vowel working space \cite{VSA_Liu_2005, Kent_2020}. These acoustic metrics have been used by speech and language pathologists (SLP) to study speech development, vowel identity and speaker characteristics in disordered speech \cite{Kent_2020}, such as after stroke \cite{Shengnan_2020}, in cerebral palsy \cite{VSA_Liu_2005} or in PD \cite{Rountrey_2020}.

In the latest studies, the vowel articulation features have been used solely, like in \cite{vanson_2018} where researchers evaluated articulation with VSA in patients with oral cancer, or together, like in \cite{ALBUQUERQUE2020} where VAI and FCR were utilized to study age and sex effects in European Portuguese vowels. As vowel articulation deficits depend much on the complexity of the speech task \cite{speech_tasks_2013}, VAI was studied in conversational spontaneous PD speech in \cite{Rountrey_2020}.

In \cite{speech_tasks_2013}, vowel articulations were studied for healthy and PD speakers in four different tasks, including sustained vowels, sentence repetition, passage reading and monologue. Beforehand, corner vowel segments were manually extracted from speech utterances. The extracted features contained the first two formants of the vowels, VSA, F2i/F2u and VAI for each speaker in different tasks. As a result, the study demonstrated that the vowel articulation indices can be used as early indicators of PD, even when speech is mildly impaired with no observable auditory degradation. Significant differences on F2u, VSA, VAI and F2i/F2u were found between people with PD and healthy groups in all speech tasks except sustained phonation. In \cite{Mou_VAI, VSA_Liu_2005}, vowel articulation indices were investigated for healthy speech and dysarthric speech related to cerebral palsy in children and adults. Significant differences in vowel acoustic indices were found between the control and cerebral palsy groups. Specifically, VSA was found to be reduced in dysarthria and significantly correlated with speech intelligibility \cite{VSA_Liu_2005}.

Currently, in clinical speech and language pathology practice, clinicians do not have automatic acoustic assessment methods for speech intelligibility. In addition, existing research on dysarthric speech has largely relied on manual annotations \cite{VSA_Liu_2005, VAI, SKODDA_2011, Sapir_2011, Skodda_2012, speech_tasks_2013, Mou_VAI}. Different vowel space measurements (e.g. VAI and VSA) are still conducted manually in the field of SLP \cite{Rountrey_2020, Caverle_2020}. Therefore, it is important to automate tools, such as VAI and VSA, to improve validity and reliability in assessing and identifying communication disorders. In addition, automatic acoustic assessment is more efficient, user-friendly and accurate compared to phonetic transcriptions or manual acoustic assessments. At best, automatic assessment has the potential in speeding up diagnostics, and the start of rehabilitation for the patients. Automatic acoustic assessment can also be used in different clinical settings, regardless of client’s age, etiology, communication disorder or functioning level.

Earlier work using ASR for pathological speech assessment has already been carried out in the context of other acoustic features.
In \cite{Zlotnik2015RandomFP, williamson2015segment}, phoneme statistics, duration and confidence measures derived from off-the-shelf Spanish ASR systems were applied to speech assessment of Spanish-speaking patients with PD. In \cite{LIU_2019}, a Cantonese ASR system was used to generate utterance-level posterior related features for broad phoneme classes in voice disorders assessment. In connection with the practical limitation that a usable ASR system may not be available for the target language, language-mismatched speech recognizer was utilized to extract phonotactic and duration features, as well as probability features in \cite{An2015Automatic}. However, to the best of our knowledge, there is no existing approach for automatic computation of the widely utilized vowel articulation features VSA, VAI, FCR or F2i/F2u. This work attempts to fill this gap by describing an automatic pipeline for vowel articulation assessment using a language-independent phoneme recognizer, followed by statistical formant analysis to account for potential errors in the recognition process.

\section{Methods}
\label{sec:method}

\subsection{System framework}

The aim of our system\footnote{Codes and scripts related to our experiments are publicly available at https://github.com/SPEECHCOG/autoVAI/} is to automate the computation of vowel articulation features in order to ease the human effort for annotation and analysis in research and clinical practice. 

\begin{figure}[htb]
    \centering
    \includegraphics[width=0.9\linewidth]{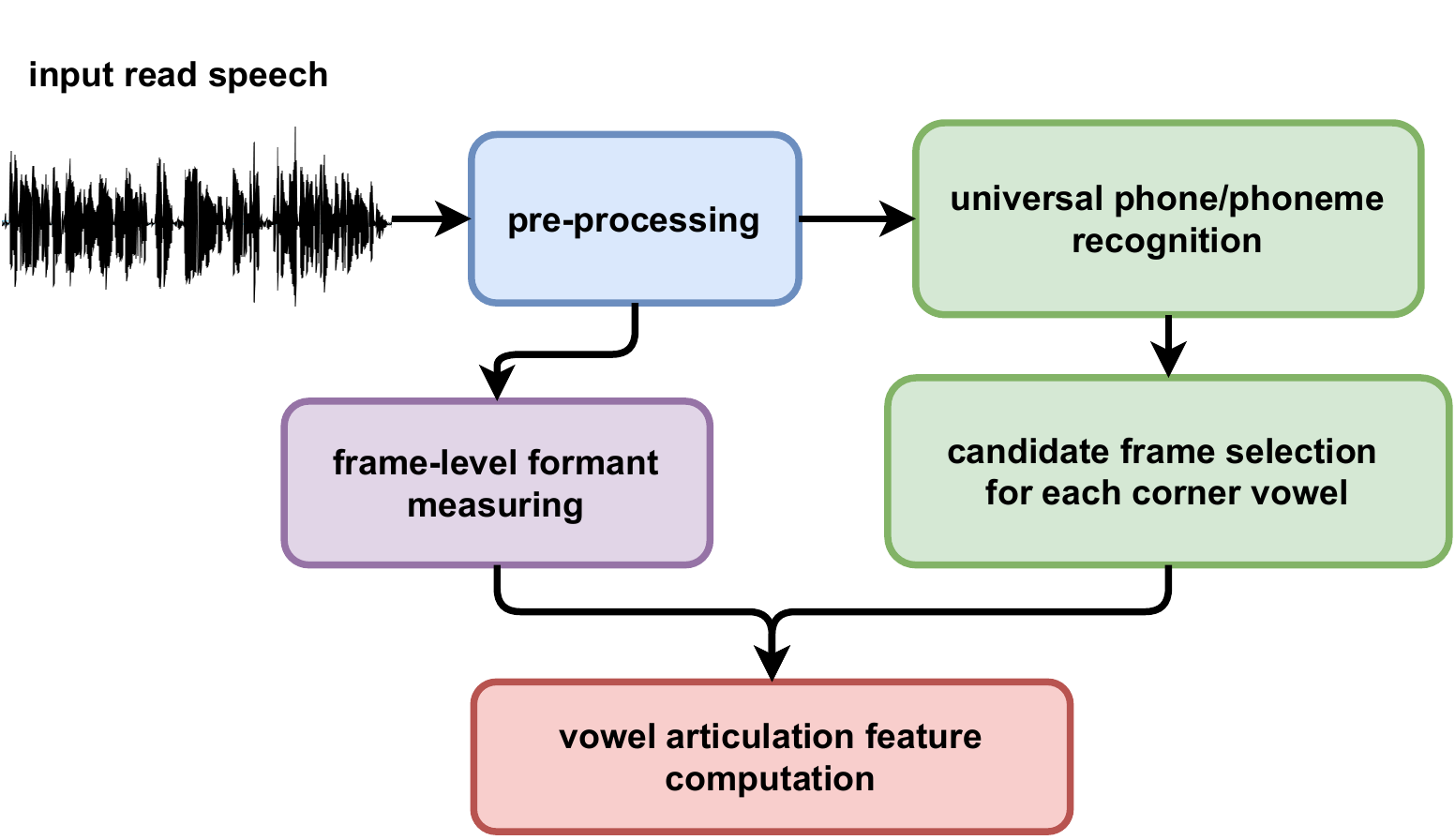}
    \caption{System framework of automatic vowel articulation feature computation.}
    \label{fig:flow}
    \vspace{-3mm}
\end{figure}

The proposed method is briefly illustrated in Fig. \ref{fig:flow}. The system aims to measure vowel articulatory undershoot in terms of four commonly utilized features, VSA, VAI, FCR and F2i/F2u, from read speech of a speaker. During pre-processing, the input speech signal is downsampled to 16 kHz. Then the read utterance is fed into a language universal phone/phoneme recognizer. Based on the recognition result, candidate frames related to each corner vowel are detected automatically. Meanwhile, formant tracking is applied to the input speech, followed by statistical analysis of the frame-level formant measurements in order to derive corner-vowel specific formant estimates. Finally, vowel articulation features are calculated from the estimates. 

\subsection{Automatic corner vowel frame detection}
\label{sub:screen}
Conventionally, in order to compute vowel articulation features, speech segments of corner vowels in chosen words are manually annotated beforehand.

In this work, an open-source universal phone/phoneme recognizer called Allosaurus \cite{ipa_recognizer} is used for corner vowel detection. It is a multi-layer Long Short-Term Memory (LSTM) neural network model trained on $12$ different languages, including English, Japanese, Mandarin, Tagalog, Turkish, Vietnamese, German, Spanish, Amharic, Italian and Russian.
It consists of a universal allophone layer with an inventory of phone units that are shared among all training languages, followed by language-dependent mappings from the %detected 
allophones to language-dependent phonemes.
In order to work with a compact and transparent phone set, the present system uses Allosaurus to obtain the most likely string of English phonemes and recognition scores (logits) for all the English allophones for a given input utterance in any language. In Allosaurus, the number of English allophones equals to the number of English phonemes, and they are from now on denoted with the ARPABET symbols \cite{Rice_1976}.

Allosaurus input frame length and hop length are set as $45$ ms and $30$ ms, respectively. For each speech frame, the phone posterior distributions are represented by a vector of logits, which can be converted to phone posterior probabilities by applying a softmax function. The logits vector has dimensionality of $208$, and its first $40$ elements that we use here correspond to English phones in the original Allosaurus training data. 

In the ARPABET notation, the corner vowels are represented as `AA' (/a/), `IY' (/i/) and `UW' (/u/).
For each of the corner vowels, there are a number of similar sounding and potentially confusable phones. For example, `AA' in `b\textbf{al}m' and `AH' in `b\textbf{u}tt' are similar in pronunciation. 
In addition, language mismatch as well as the acoustic differences between recognizer training and usage data can contribute to recognition confusion. Notably, given the text-free nature of our system, it is also likely that severely dysarthric speakers may fail to reach the articulatory targets of corner vowels, resulting in a more centralized phone production instead. In this case, the recognizer may recognize the sound as a centralized phone according to its acoustic properties instead of the original intended sound by the speaker. However, for our corner vowel articulation analysis, it is important to capture these productions as exemplars of corner vowels as well. Considering the likely recognition confusion, we extended each individual corner vowel to a set of potentially confusable phones/phonemes. Table \ref{tab:phone_sets} shows the list of Allosaurus English phone categories that we associate with each corner vowel. The benefits of using extended phone/phoneme sets will be discussed in Section \ref{sub:disc_wide_phones}.

\begin{table}[!htbp]
    \centering
    \caption{Phone/phoneme sets for corner vowels.}
    \begin{tabular}{ll}
        \hline
         Corner vowel & Related phones/phonemes \\
         \hline
         $\textbf{Z}_a$: /a/ & `AA', `AE', `AH', `AW', `AY' \\
         $\textbf{Z}_i$: /i/ & `IY', `IX', `IH' \\
         $\textbf{Z}_u$: /u/ & `UW', `UH', `OW' \\
        \hline
    \end{tabular}

    \label{tab:phone_sets}
\end{table}

To screen out the frames which are most likely to be recognized as corner vowels, two selection criteria were designed. One is based on recognition result of the most likely phoneme sequence while the other is based on frame-level phone posterior probabilities. The screening process is described as following.
First the input utterance $\textbf{X}=\{\textbf{x}_t\}$ is decoded by Allosaurus. From the output of Allosaurus, recognition of the most likely phoneme sequence $\{z_t\}$ together with logits vectors $\textbf{Y}=\{\textbf{y}_t\}$ are obtained. To obtain the English phone posteriors $\textbf{P}=\{\textbf{p}_t\}$, the first $40$ dimensions in $\textbf{y}_t$ related to English model are kept while the others are discarded, followed by softmax function. The elements in $\textbf{p}_t$ indicate how likely a frame at time \textit{t} corresponds to each phone. Briefly speaking, the two selection criteria are described as:
\begin{enumerate}
    \item Frame \textit{t} will be selected as a candidate of corner vowel if $z_t$ belongs to one of the phoneme sets listed in Table \ref{tab:phone_sets}.
    \item Frame \textit{t} can also be selected according to its phone posterior distribution, if among the top \textit{k} phones with the highest posterior probabilities in $\textbf{p}_t$, any of them lies in the corner vowel phoneme sets and has posterior larger than a predefined threshold $\alpha$. 
\end{enumerate}
The present system uses $k=4$ and $\alpha=0.2$ based on qualitative observations. Theoretically, one frame could be counted for more than one corner vowels. After the automatic screening, three frame sets $\textbf{S}_a$, $\textbf{S}_i$ and $\textbf{S}_u$ are generated for the given utterance, corresponding to /a/, /i/ and /u/, respectively.

\begin{figure*}[htb]
    \centering
    \includegraphics[width=0.8\linewidth]{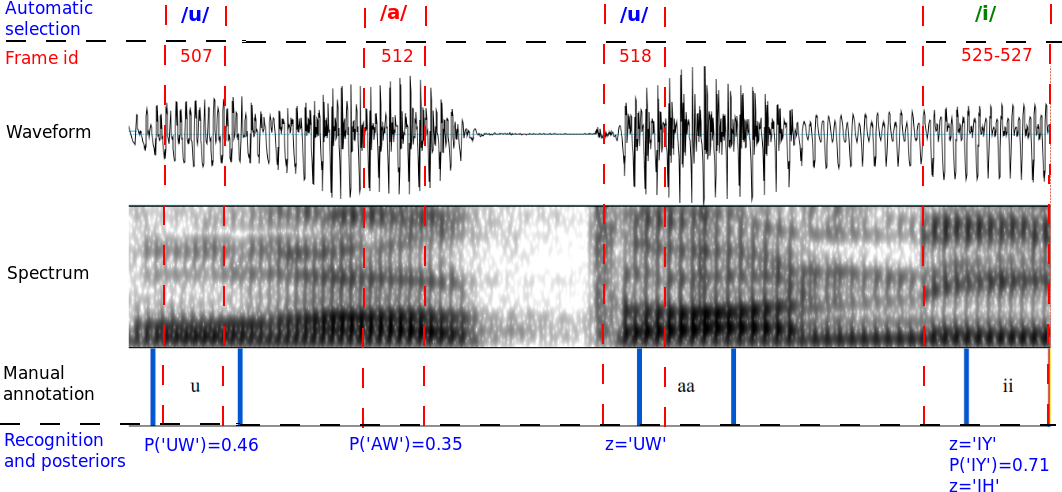}
    \caption{Example of automatic corner-vowel-related frames selection (utterance of Finnish words `P\textbf{u}halt\textbf{aa} n\textbf{ii}n'). As a reference, manually annotated stable phonation sections of pre-selected vowels are also shown (see section \ref{sub:finnish} for details).}
    \label{fig:phn_post_L12}
    \vspace{-3mm} 
\end{figure*}

An example of automatic frames selection is illustrated in Fig. \ref{fig:phn_post_L12}. The plots show the waveform, spectrum, manual annotation of stable center sections of corner vowels together with the most likely recognition output and dominant phone posteriors for approximately $20$ frames. The linguistic content of this segment is a Finnish utterance `P\textbf{u}halt\textbf{aa} n\textbf{ii}n' (pronunciation in IPA, [p u h a l t a:] [n i: n]), where center stable segments  /u/, /a:/, and /i:/ were manually marked. Frame $507$ was manually annotated as /u/. And it was automatically selected for /u/, since it got a posterior of $0.46$ for `UW' (lying in $\textbf{Z}_u$). Similarly, frame $512$ was selected for /a/, which corresponds to the sound of `a' in `Puh\textbf{a}ltaa'. Unfortunately, frame $518$, with overlap of sound `aa', was wrongly recognized as `UW' related to /u/. Frames $525-527$ were manually annotated as `ii' and automatically detected as /i/ based on the recognized phonemes (`IY' and `IH') and phone posterior ($0.71$ for `IY'). The efficacy of automatic corner-vowel-related frames selection will be discussed in Section \ref{sub:disc_frames_selection}.

\subsection{Corner vowel representation}
\label{sub:vowel_representation}

Based on the automatic corner-vowel-related frames selection introduced above, three frame groups $\textbf{S}_a$, $\textbf{S}_i$ and $\textbf{S}_u$ related to /a/, /i/ and /u/ are obtained for the input speech. Any selected frame $\textbf{x}_t$ is then represented by its first two formants, which were estimated with Burg's algorithm \cite{burg}.

The next step is to obtain a single estimate for the corner vowel formant frequencies to be used in vowel articulation feature equations. In order to do this, the frame-level estimates of F1 and F2 of speech corresponding to each corner vowel are averaged. This is to align with what the speech therapists (also the annotators in this work) do to measure vowel articulation, where they compute average formant frequencies across manually segmented vowels. 
The process is repeated for all frames in $\textbf{S}_a$, $\textbf{S}_i$ and $\textbf{S}_u$, resulting in vowel-specific representative formant estimates ($\rm F1a$, $\rm F2a$), ($\rm F1i$, $\rm F2i$) and ($\rm F1u$, $\rm F2u$), respectively. However, the automatic frames selection has a known tendency for centralized estimates and may also suffer from outliers due to recognition errors, which may bias the mean estimates in an undesirable manner. In order to reduce the impact of the centralized frames and to increase the corner vowel representativeness of the detected frames, we explored the use of 70/30 and 90/10 percentiles for high/low formants, respectively, and median (50/50 percentile) in addition to the mean estimates. For example, in the 70/30 percentiles case, the formant estimate of /a/ is the $70$th percentile of F1 and the $30$th percentile of F2 among selected /a/ frames, whereas vowel /i/ is represented by the $30$th and $70$th percentiles of F1 and F2, respectively. Vowel /u/ is represented by the $30$th percentile of both F1 and F2. 

\begin{figure*}[htb]
\centering
\subfigure[automatic selection]{
\includegraphics[width=0.45\linewidth]{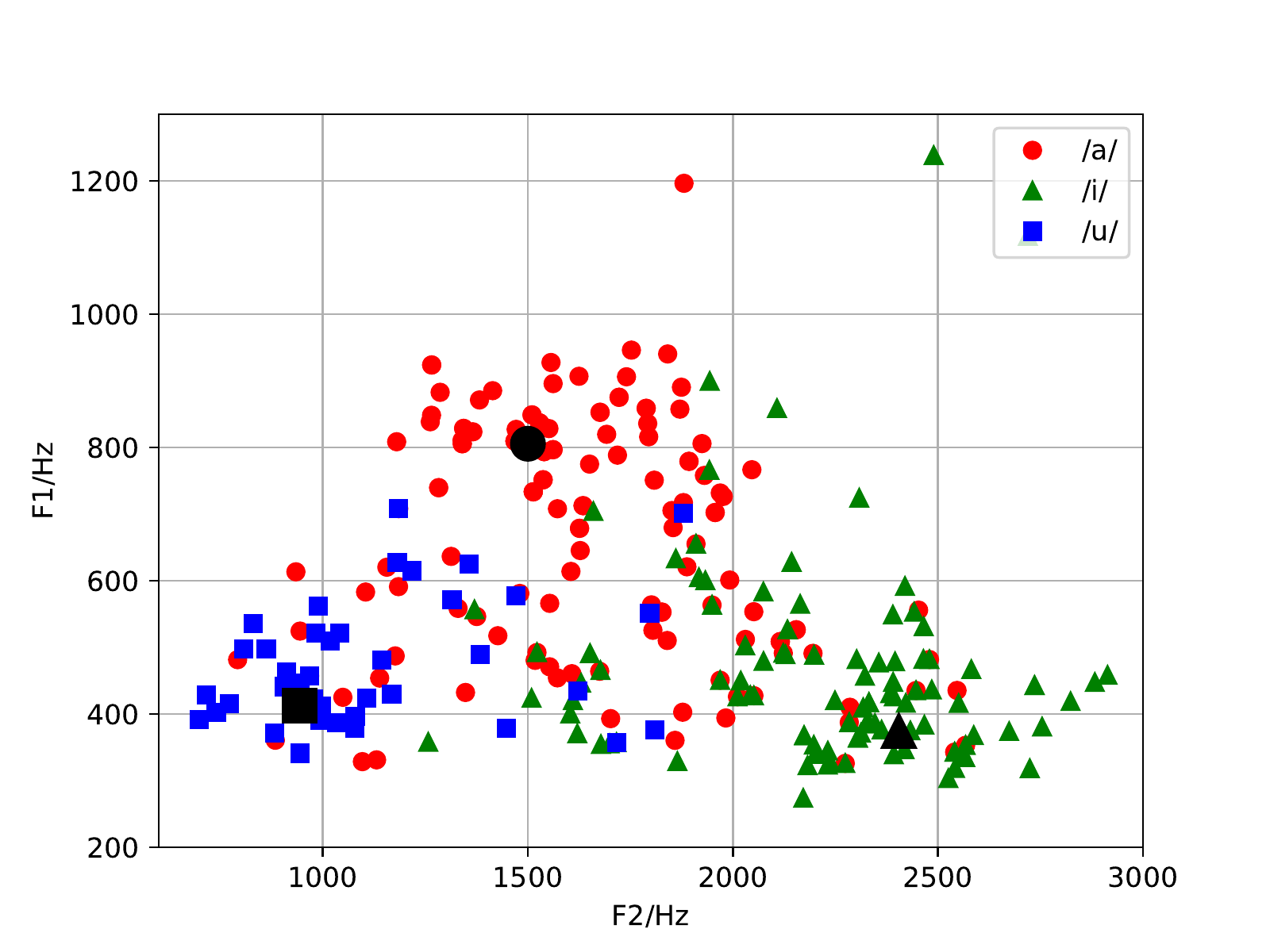}
}
\subfigure[manual annotation]{
\includegraphics[width=0.45\linewidth]{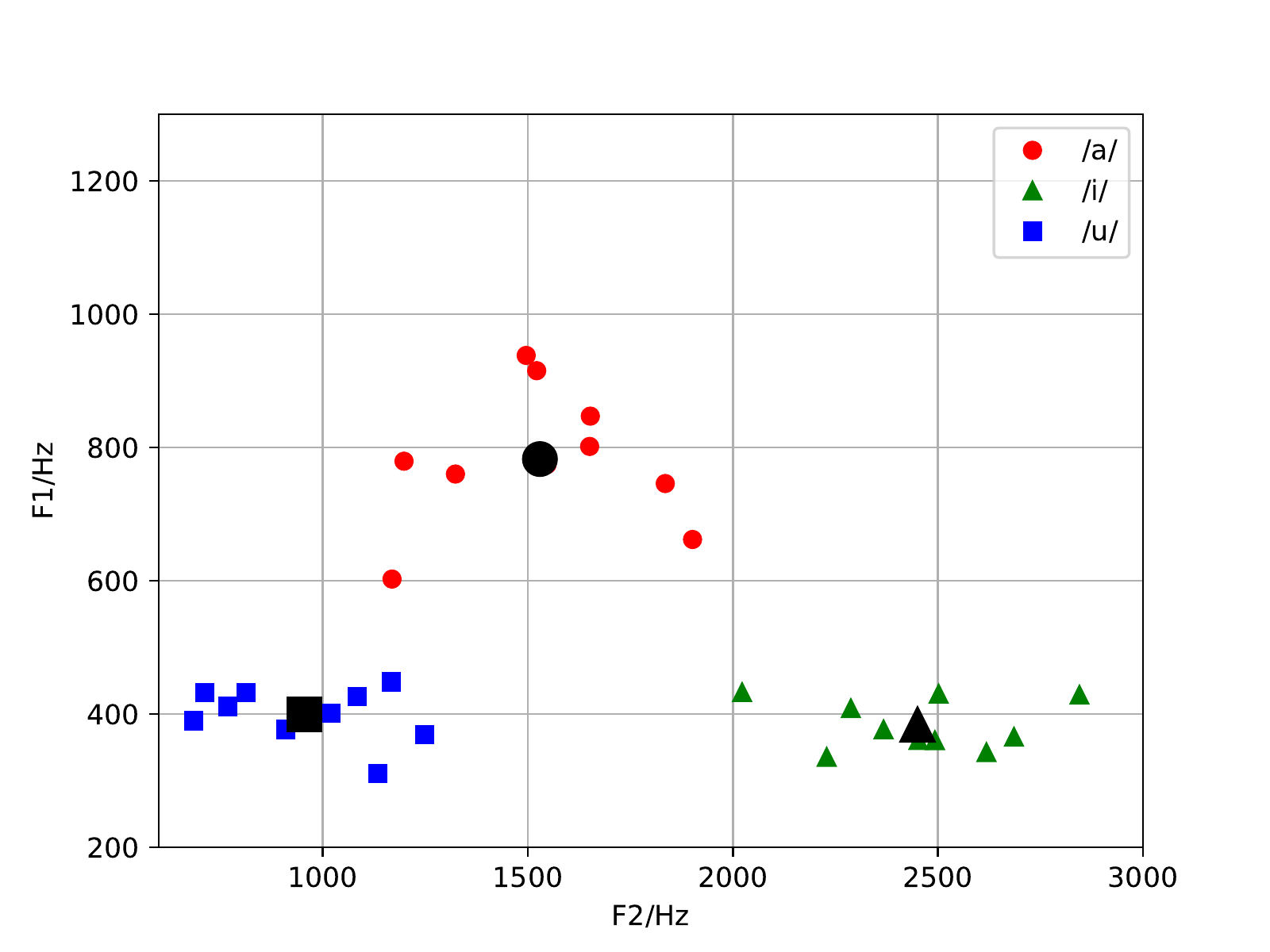}
}
\caption{Corner vowel frame/segment distributions automatically/manually selected from one Finnish speaker's speech. On the left, each colored point corresponds to an estimate for one signal frame. On the right, each colored point stands for an estimate for an annotated vowel segment. The solid black symbols are the representative estimates of the corner vowels. In the automatic selection, corner vowels are represented by 70/30 percentiles of formants of selected frames (introduced in Section \ref{sub:vowel_representation}). In manual annotation, corner vowels are represented by the mean formants of the annotated segments.}
\label{fig:frames_K7}
\vspace{-3mm} 
\end{figure*}

An example of automatic corner vowel representation with automatic selection is shown in Fig. \ref{fig:frames_K7}. Formant estimates based on manual vowel center segment annotation are shown in the right panel for comparison. For both methods, the relative positions of three corner vowels are similar, where /a/ tends to have a high F1 while /i/ tends to have a high F2 and /u/ has low values for F1 and F2. However, as expected due to the broad phoneme grouping (Table \ref{tab:phone_sets}), there are also automatically detected frames distributed towards the center of vowel triangle. In Fig. \ref{fig:frames_K7}, representative formant values (from now on `representatives') of selected frames of each corner vowel are shown by black solid symbols (dots, squares and triangles). In the automatic method, representatives were calculated with 70/30 percentiles of formants while those in manual method were represented by mean formants of annotated segments. As shown in this figure, the representatives for each vowel are located almost in the same position for the two methods. 

\subsection{Vowel articulation}
\label{sub:vowel_articulation}
The most commonly used features to quantify the vowel articulation undershoot in both clinical and acoustical research include VSA, VAI, FCR and F2i/F2u, which are computed from F1 and F2 of corner vowels. Here corner vowels /a/, /i/ and /u/ are represented by the formant estimates described in the previous subsection, i.e. ($\rm F1a$, $\rm F2a$), ($\rm F1i$, $\rm F2i$) and ($\rm F1u$, $\rm F2u$) respectively.
\begin{itemize}
    \item \textbf{VSA} is computed as the area of a triangle formed with vertices of corner vowels in the F1-F2 plane (Eq. \ref{eq:vsa}). VSA is expected to be compressed with formant centralization \cite{Kent_2003}.
    \begin{equation}
    \setlength{\abovedisplayskip}{1pt}
    \setlength{\belowdisplayskip}{1pt}
        \begin{split}
        \rm VSA = \frac{1}{2}|F1i(F2a-F2u) + F1a(F2u-F2i) + \\ \rm F1u(F2i-F2a)|
        \label{eq:vsa}
        \end{split}
    \end{equation}
    \item \textbf{VAI} is expressed as a ratio of corner vowels' formants (Eq. \ref{eq:VAI}). VAI has been proven to be more robust to inter-speaker variability than VSA \cite{VAI}. With formant centralization, the numerator of VAI is likely to decrease while the denominator is likely to increase.
    \begin{equation}
        \setlength{\abovedisplayskip}{1pt}
    \setlength{\belowdisplayskip}{1pt}
        \rm VAI = \frac{F1a + F2i}{F2a + F1i + F1u + F2u}
        \label{eq:VAI}
    \end{equation}

    \item \textbf{FCR} is the reciprocal value of VAI, which was designed to increase with vowel centralization and decrease with vowel expansion \cite{FCR}.
    \item \textbf{F2i/F2u} is the ratio of F2 of vowel /i/ and /u/, which was also demonstrated to be less sensitive to inter-speaker variability in \cite{FCR}.
\end{itemize}

Vowel articulations computed with the hi/lo percentiles of formants are symbolized by $\rm VAI[hi]$, $\rm VSA[hi]$, $\rm FCR[hi]$ and $\rm F2i/F2u[hi]$ respectively. For example, 
\begin{equation}
    \setlength{\abovedisplayskip}{1pt}
    \setlength{\belowdisplayskip}{1pt}
    \rm VAI[hi] = \frac{F1a[hi] + F2i[hi]}{F2a[lo] + F1i[lo] + F1u[lo] + F2u[lo]}
\end{equation}
The logic is similar for computation of $\rm VSA[hi]$, $\rm FCR[hi]$ and $\rm F2i/F2u[hi]$ using percentiles formants. Here $\rm hi + lo = 100$ and $\rm hi \ge lo$. In our experiments, $\rm hi$ was set to $50$, $70$ and $90$ for investigation. Accordingly, $\rm lo$ equaled to $50$, $30$ and $10$.

\section{Speech corpora}
\label{sec:corpus}

\subsection{Finnish PD speech corpus}
\label{sub:finnish}
In order to validate our automated system for vowel articulatory undershoot quantification, a subset of Finnish read speech from the newly collected Parkinson's Disease Speech corpus of Tampere University (PDSTU) was used. The subset contains reading speech from $67$ native Finnish speakers, including $35$ speakers diagnosed as PD as well as $32$ control speakers. The speech were recorded with a close-talking microphone as mono channel with 32 bits and  with a sampling rate of 44.1 kHz at the Tampere University. 
The reading material was a passage ``\textit{Pohjantuuli ja aurinko}'' (``\textit{North Wind and the Sun}''),
containing $77$ Finnish words as listed below. This reading sample has been commonly used in clinical and research settings in Finland (e.g. in \cite{Kankare_2017}) and it is comparable to the Rainbow passage often used in English studies. 

\begin{itemize}

    \item \textbf{Read passage:}

    Pohj\textbf{a}nt\textbf{uu}li ja aur\textbf{i}nko väittel\textbf{i}vät, k\textbf{u}mm\textbf{a}lla ol\textbf{i}si enemmän voim\textbf{aa}, kun he s\textbf{a}malla näk\textbf{i}vät kulkij\textbf{a}n, jolla oli yllään lämmin t\textbf{a}kki. Silloin he sopivat, että se on voimakk\textbf{aa}mpi, joka nopeammin saa k\textbf{u}lkijan r\textbf{ii}s\textbf{u}m\textbf{aa}n takk\textbf{i}nsa. Pohjant\textbf{uu}li alkoi p\textbf{u}halt\textbf{aa} n\textbf{ii}n, että viuhui, mutta mitä kovemp\textbf{aa} se p\textbf{u}halsi, sitä tarkemmin kääri mies takin ympärilleen, ja v\textbf{ii}mein t\textbf{uu}li luopui koko hommasta. Silloin alkoi aurinko loistaa lämpimästi, eikä aikaakaan, niin kulkija r\textbf{ii}sui manttelinsa. N\textbf{ii}n oli t\textbf{uu}len pakko myöntää, että aurinko oli kuin olikin heistä vahvempi.

\end{itemize}

The data were manually annotated according to clinical practice by three speech researchers (authors N.P., T.I. and J.L.), using both auditory perception and visual spectral analysis in Praat. The phonemes marked in bold are the corner vowels that were chosen for manual annotation, consisting of $5$ \textbf{aa} and $5$ \textbf{a} for /a/, $5$ \textbf{ii} and $5$ \textbf{i} for /i/, $4$ \textbf{uu} and $6$ \textbf{u} for /u/. In Finnish, vowels have two durations (quantities), short and long. The short vowel is written with one letter while the long vowel is written with two equal letters. For each speaker, a minimum of $30$ ms segment from a stable phonation at the temporal midpoint of each chosen vowel was manually segmented and annotated for the first two formants for vowel articulation measurement. The process is different from the regular phonetic transcription which marks the beginning and ending of a phoneme's pronunciation. After removing leading and trailing silences from speech of reading the passage, the average length of reading speech per each PD speaker was approximately $40.0$ s while the minimum and maximum of speech duration were $25.7$ s and $55.1$ s, respectively. For control speakers, the average, minimum and maximum reading times were $34.8$ s, $28.5$ s and $40.9$ s, respectively.

Besides the manual annotation, the read speech from the $35$ PD speakers and a random sample of $15$ control speakers was rated by $3$ external experts who had an average $23$ years (and at least $16$ years) of working experience in speech therapy, and who were recruited through public advertisement. The rated dimensions include speech intelligibility, voice impairment and overall severity of communication disorder according to their standard definitions in SLP: \textit{Speech intelligibility} refers to the degree to which a spoken utterance is understood by a listener. \textit{An impaired voice} may be characterized by altered vocal quality, pitch, loudness, or vocal effort. Vocal quality factors include but are not limited to: roughness, breathiness, strained quality, hoarseness, and tremulous voice. \textit{A communication disorder} is a general impairment in the ability to effectively receive, send, process, and comprehend spoken information \cite{disordersdef}.

For speech intelligibility, a standard sample was selected from a healthy control speaker and its intelligibility was defined as $100$. Each rater was asked to compare the intelligibility of the recording to be rated with that of the standard sample. Scores larger (smaller) than $100$ means more (less) intelligible than the standard sample. When rating the intelligibility, three short randomly selected phrases from the reading passage were presented. The samples selected for each speaker were different to reduce familiarization. The rater could only listen to the presented sample once. The ratings of voice impairment and overall severity of communication disorder were carried out using a scale from 0 (normal) to 100 (most severe). To rate voice impairment and overall severity, the raters were allowed to listen to the presented samples as many times as they needed. The recordings were presented in a randomized order and participant information was hidden to the raters. The test was conducted in a quiet room using high-quality headphones (Sennheiser HD598). 

Severity of the PD was assessed using Hoehn and Yahr scale (H\&Y), obtained using a dedicated questionnaire administered to the PD subjects. The scale has $8$ values from $1$ to $5$ in $0.5$ increments and describes the severity of movement disorder \cite{HY}. The larger the value, the severer the movement disorder.

Statistics of the dataset are summarized in Table \ref{tab:spkinfo_pdstu}, where the numbers denote the mean and the minimum and maximum of each value inside parentheses.
The mean H\&Y of $1.8$ indicates, on average, a mild stage of PD in our subject population. However, symptoms of PD on speech and voice can be already seen from the expert ratings, where average PD patient intelligibility is lower and impairment and disorder ratings higher than those of controls.

Permission to conduct the present study on PDSTU was obtained from the Ethics Committee of Tampere University. All subjects provided a written informed consent according to the Declaration of Helsinki.

\begin{table}[]
    \centering
    \caption{Participant statistics for the Finnish PDSTU corpus (mean, (min, max)). $^{\dag}$Experts' ratings for control group were based on recordings from a sample of $15$ control speakers.}
    \label{tab:spkinfo_pdstu}
    \begin{tabular}{lll}
    \hline
         &\textbf{ Control speakers }& \textbf{PD speakers} \\
         \hline
        Female & 20 & 21 \\
        Male & 12 & 14 \\
        Age & 49.9 (24, 67) & 65.6 (48, 82) \\
        Years after diagnosis & N/A & 5.5 (1, 18) \\
        H\&Y & N/A & 1.8 (1, 2.5) \\
        Speech intelligibility & 97.3 (76.0, 114.2)$^{\dag}$ & 78.3 (45.0, 101.7) \\
        Voice impairment & 12.4 (2.5, 26.1)$^{\dag}$ & 28.3 (5.7, 75.2) \\
        Overall disorder & 9.5 (1.2, 20.8)$^{\dag}$ & 23.2 (4.0, 66.9) \\
    \hline
    \end{tabular}
\end{table}

\subsection{Spanish PD speech corpus}
\label{sub:pcgita}

In order to further validate the reliability and language-independence of our proposed method for automatic vowel articulation feature computation, two more speech corpora in different languages were used in this work.  

First of them, PC-GITA, is a widely used Spanish PD speech database \cite{PCGITA}, which contains speech recordings from $50$ speakers diagnosed with PD and $50$ control speakers matched by age and gender. All the speakers are native Spanish speakers. In this work, only text reading speech were used for experiments. For the PD speakers, the average length of reading sample was approx. $18.6$ s, with minimum and maximum durations of $28.5$ s and $40.9$ s. For the control speakers, the average, minimum and maximum sample lengths were $17.6$ s, $11.2$ s and $34.0$ s, respectively. 
Together with the audio recordings, participants' information for both control and PD groups were provided, shown as Table \ref{tab:spk_pcgita}. The larger the UPDRS (UPDRS-speech), the severer the PD (speech impairment).
\begin{table}[!htbp]
    \centering
    \caption{Statistics of participant information in the Spanish corpus, PC-GITA (mean, (min,  max)).}
    \label{tab:spk_pcgita}
    \begin{tabular}{llll}

    \hline
         & Male & Female & Total \\
    \hline
    \textbf{Control speakers} & 25 & 25 & 50\\
    Age & 60.5 (31, 86) & 61.4 (49, 76) & 61.0 \\
    \hline
    \textbf{PD speakers} & 25 & 25 & 50 \\
    Age & 61.3 (33, 81) & 60.7 (49, 75) & 61.0 \\
    Years after diagnosis & 8.7 (0.4, 20) & 13.8 (1, 43) & 11.2 \\
    H\&Y & 2.1 (1, 4) & 2.2 (1, 3) & 2.2 \\
    UPDRS & 37.8 (6, 93) & 37.6 (19, 71) & 37.7 \\
    UPDRS-speech & 1.4 (0, 3) & 1.3 (0, 3) & 1.3 \\
    \hline
    \end{tabular}
\end{table}

\subsection{English dysarthric speech corpus}
\label{sub:torgo}
TORGO \cite{TORGO}, a widely used English corpus, contains speech from both control group and dysarthric speakers with either cerebral palsy or amyotrophic lateral sclerosis. The data are free to download from the TORGO website \footnote{http://www.cs.toronto.edu/~complingweb/data/TORGO/torgo.html}, consisting of $7$ control speakers ($3$ females and $4$ males) and $8$ dysarthric ($3$ females and $5$ males) speakers. The dysarthric speakers were between the ages of $16$ to $50$ years old and the control speakers were age-matched. 
Speech tasks in TORGO include non-words, short-words and restricted/unrestricted sentences. In order to make the data comparable to a passage reading task, we concatenated multiple short sentences together. For each session of a speaker, the $20$ sentences with the longest prompts were selected. Then the corresponding $20$ speech recordings were utilized to form two long waveforms, each of which contains ten sentences. As a result, a total of $52$ long waveforms from the $15$ speakers were obtained (not all sessions were available from all talkers). Among the $52$ signals, half were from the control group and half from the dysarthric group. The mean, maximum and minimum of durations of these audio files were approx. $70$ s, $137$ s and $42$ s. All selected signals were recorded with a head-mounted microphone. The speaking content varied across the concatenated audios.
Following \cite{Joy_torgo_2018}, the $8$ dysarthric speakers were divided into $3$ severity levels according to overall clinical intelligibility and articulatory function, with $2$ mildly, $2$ moderately and $4$ severely disordered speakers.
For correlation analyses, we assigned a numerical severity score of `0, 1, 2, 3' to the `control, mild, moderate, severe' speakers, respectively.

\section{Experiments}
\label{sec:result}
Based on the manually annotated corner vowel segments, gold-standard VAI, VSA, FCR and F2i/F2u features were computed for each speaker in the Finnish corpus PDSTU. In order to validate our system, the automatic vowel articulation measurement system was compared against the same measures based on hand-annotated speech segments for the Finnish dataset.

\subsection{Consistency of automatic vowel articulation features}

Table \ref{tab:corr_auto_man} lists the Pearson correlation coefficients between the automatic and manual computations on various dimensions on $67$ speakers of the Finnish corpus. The vowel articulation features include VAI, VSA, FCR and F2i/F2u computed with the mean and different percentiles (hi/lo) of frame-level formants as defined in Section \ref{sub:vowel_articulation}. Besides the vowel articulation features, correlations on corner vowel formant estimates are also reported. All the correlation coefficients are larger than $0.73$ while the largest correlation values for each feature are in range of $0.81 - 0.97$, marked as \textbf{bold} in the table ($p < .00001$ for all measurements). On average, the correlations in the control group and in the PD group are similar.
All coefficients in Table \ref{tab:corr_auto_man} show strong and significant correlation for features computed with manual and automatic methods. Therefore, the result indicates the reliability of the proposed automatic approach for vowel articulation feature computation. In addition, the use of percentiles can improve the correlation coefficients over the mean F2a, F1i, F1u and F2u when computing the features.

\begin{figure*}[htb]
\centering
\subfigure[VAI]{
\includegraphics[width=0.45\linewidth]{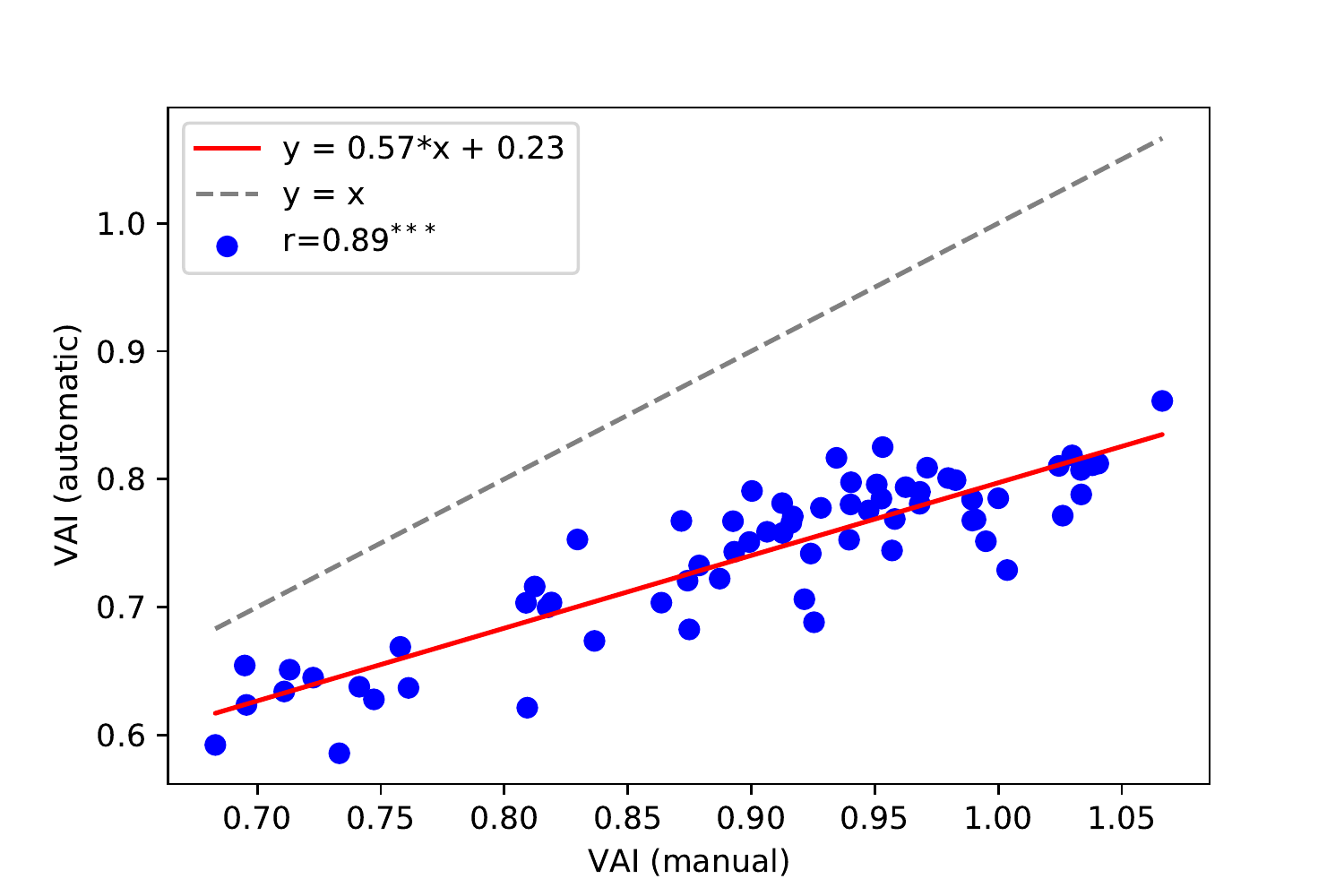}
}
\vspace{-3mm}
\subfigure[VSA]{
\includegraphics[width=0.45\linewidth]{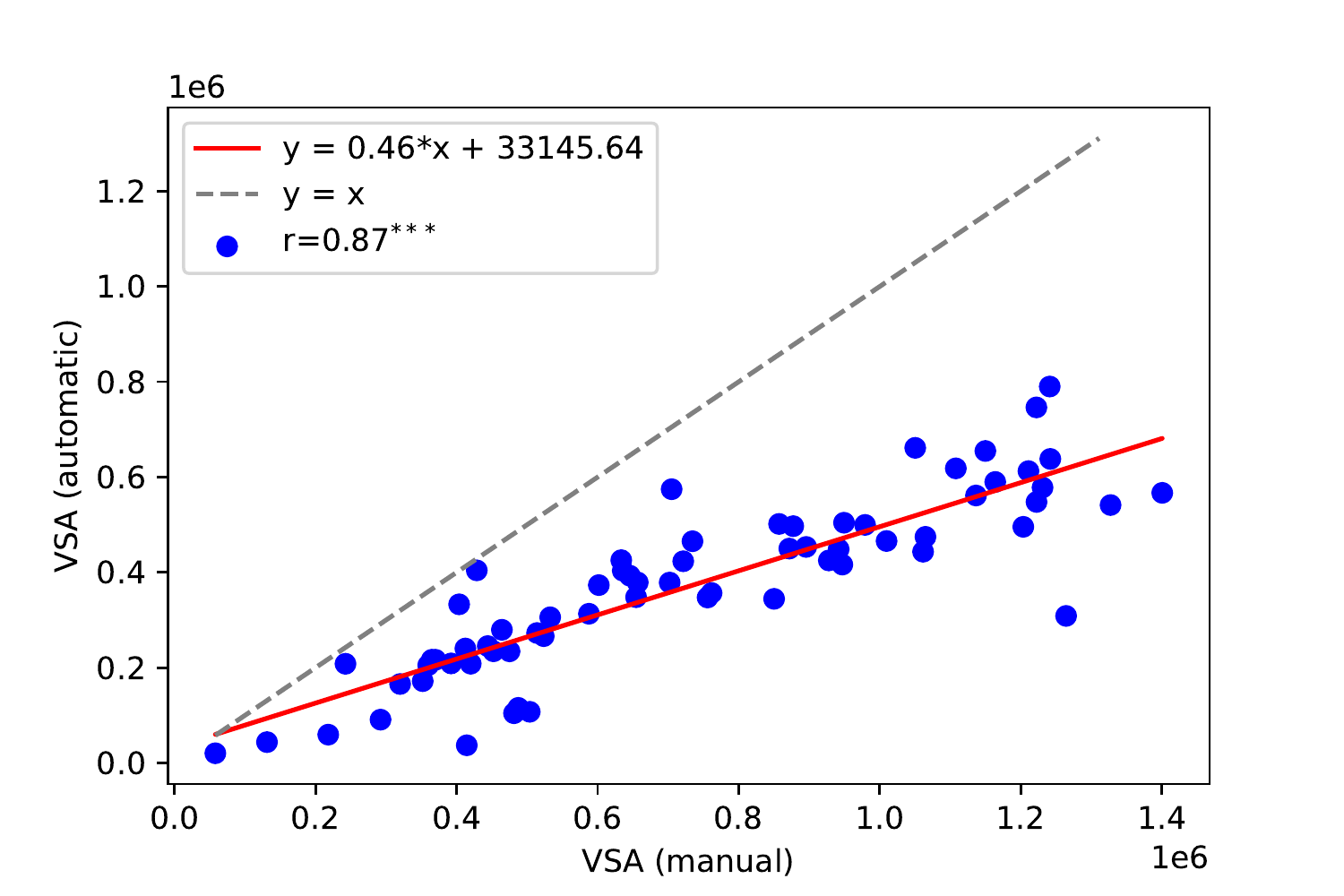}
}
\vspace{-3mm}
\quad

\subfigure[FCR]{
\includegraphics[width=0.45\linewidth]{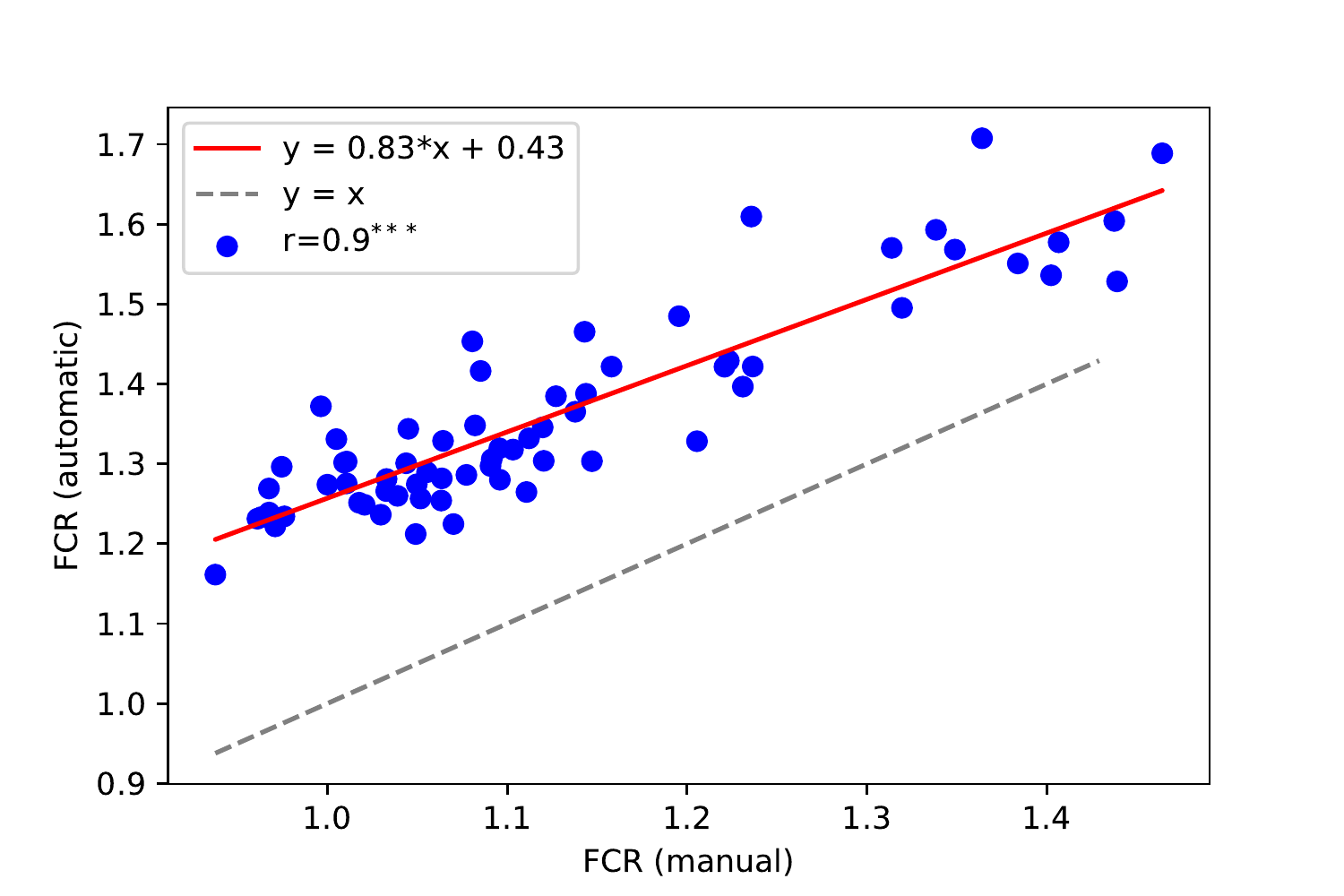}
}
\subfigure[F2i/F2u]{
\includegraphics[width=0.45\linewidth]{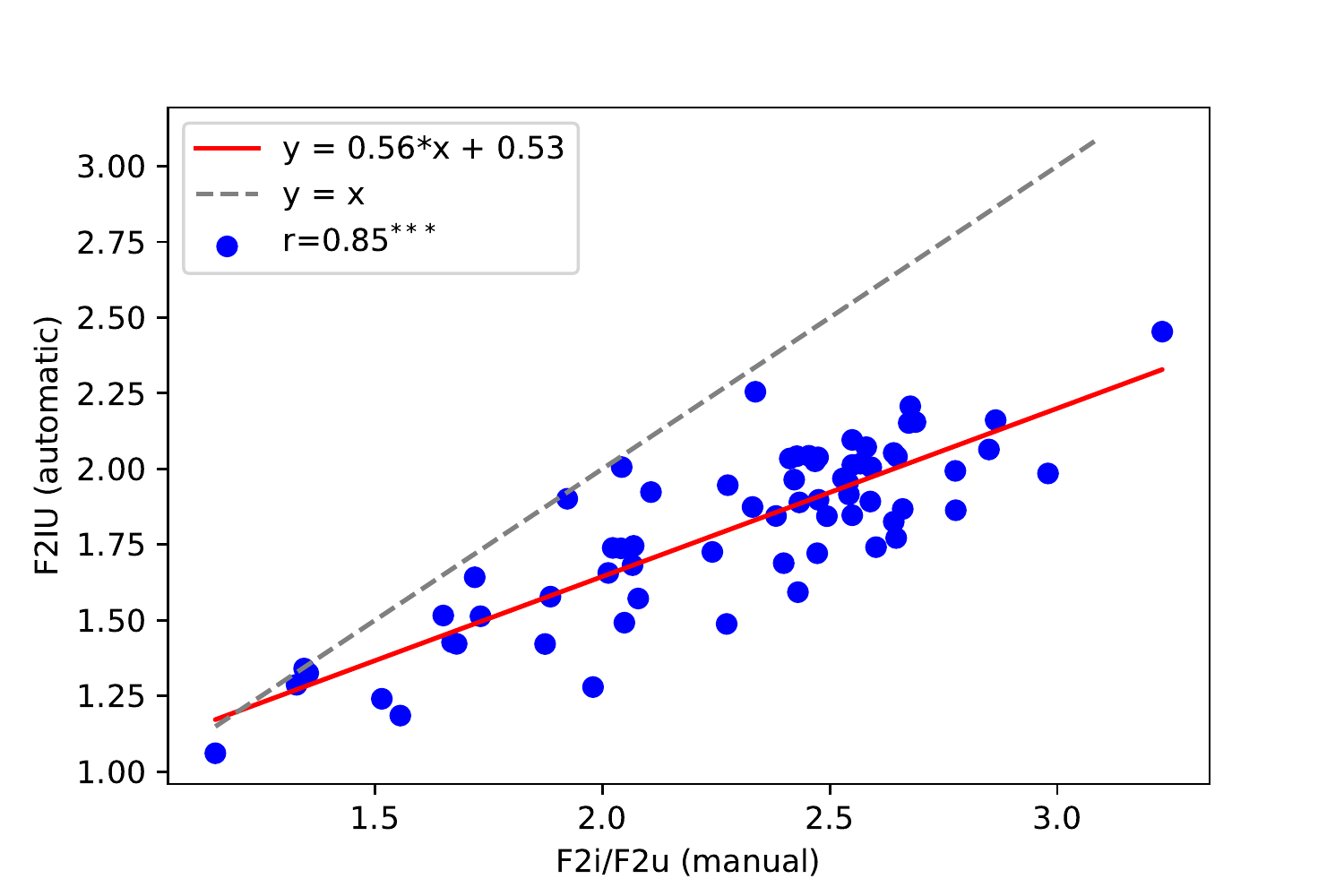}
}
\caption{Correlations of vowel articulation features computed with the mean of formants under automatic selection and manual annotation for the 67 speakers in the Finnish speech corpus PDSTU. In the legend, `r' stands for Pearson correlation coefficient and three-stars ($^{***}$) indicates statistical significance at $p < .001$.}
\label{fig:corr_man_auto}
\vspace{-3mm}
\end{figure*}

\begin{figure*}[htb]
\centering
\subfigure[VAI]{
\includegraphics[width=0.45\linewidth]{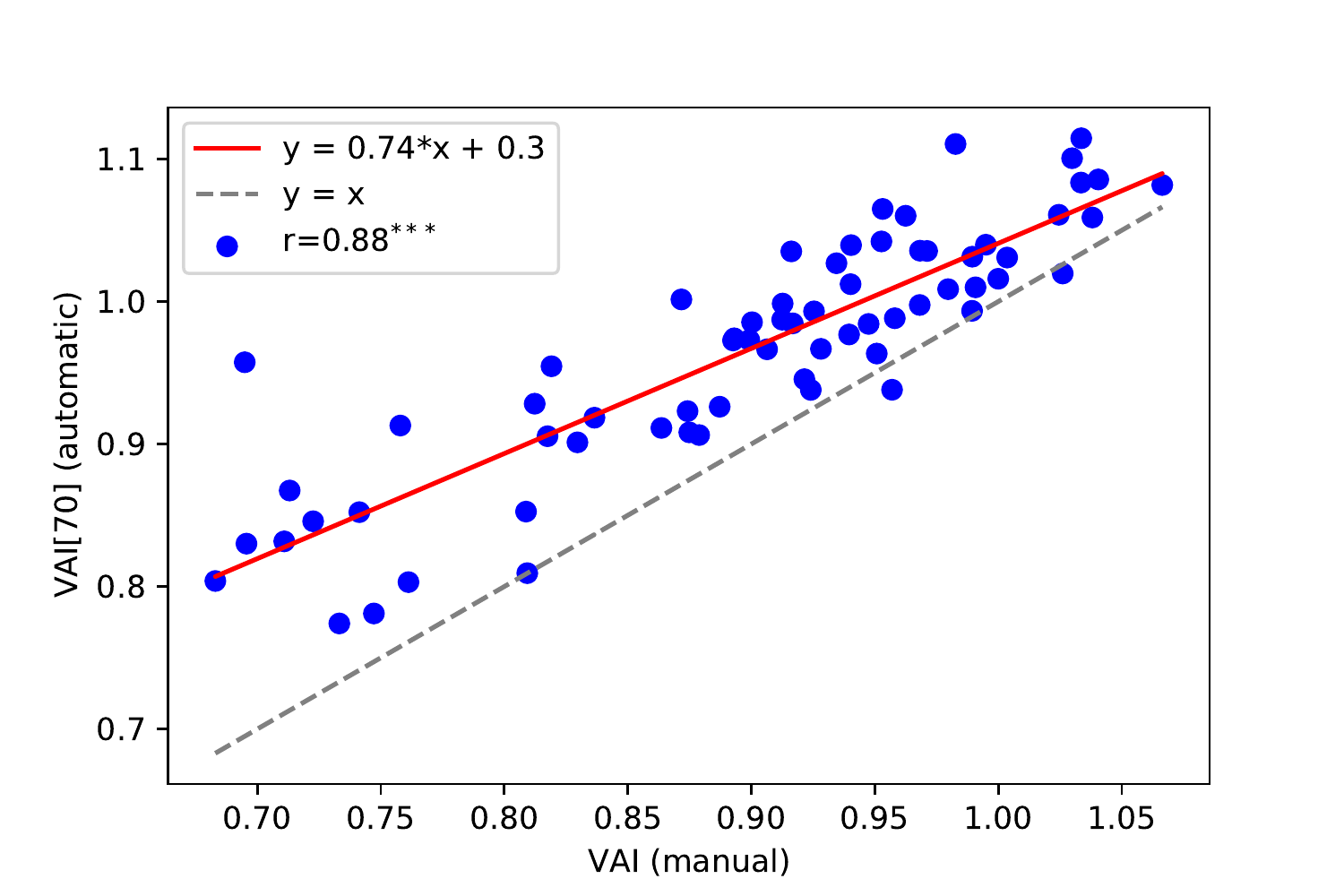}
}
\vspace{-1mm}
\subfigure[VSA]{
\includegraphics[width=0.45\linewidth]{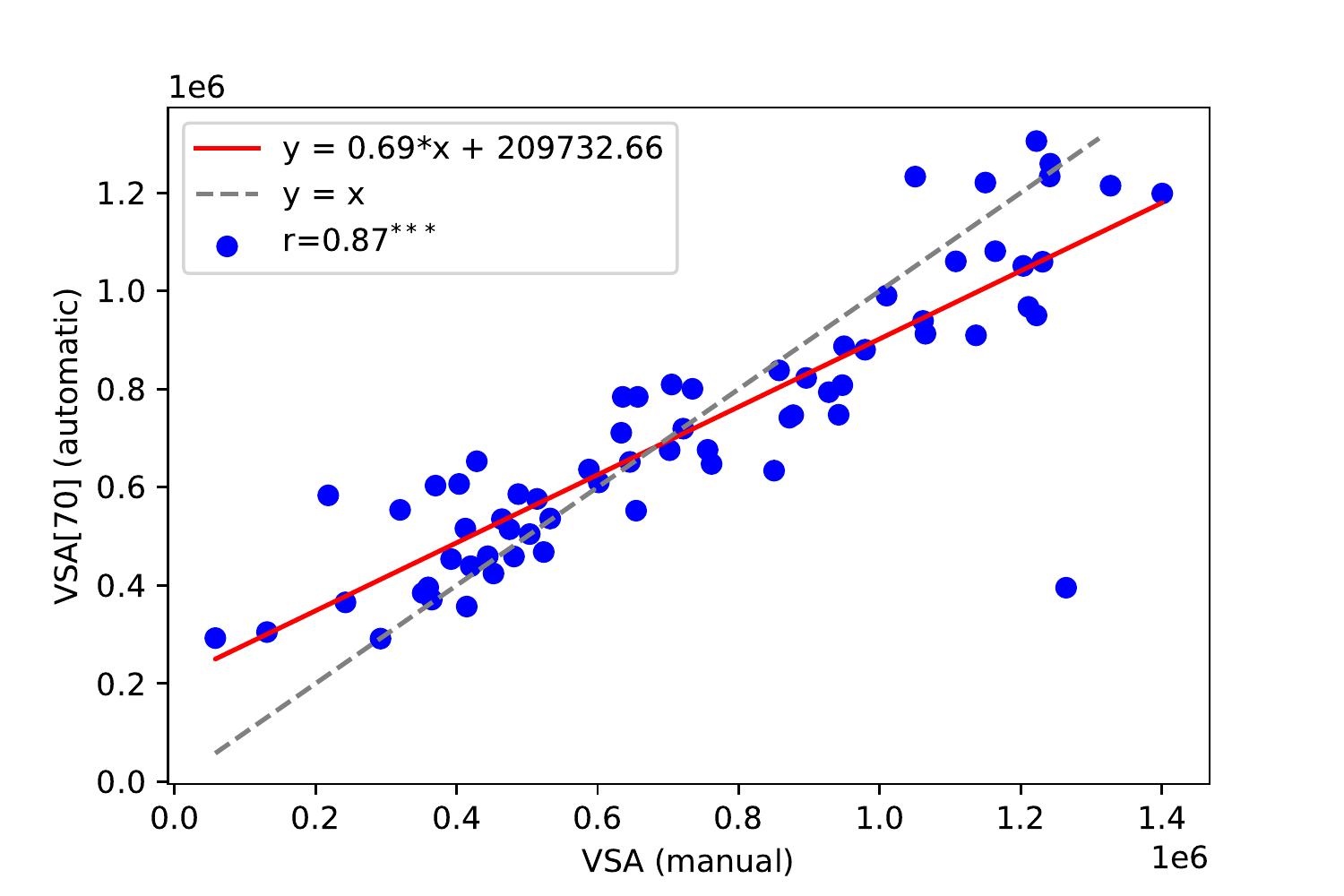}
}
\vspace{-1mm}
\quad
\subfigure[FCR]{
\includegraphics[width=0.45\linewidth]{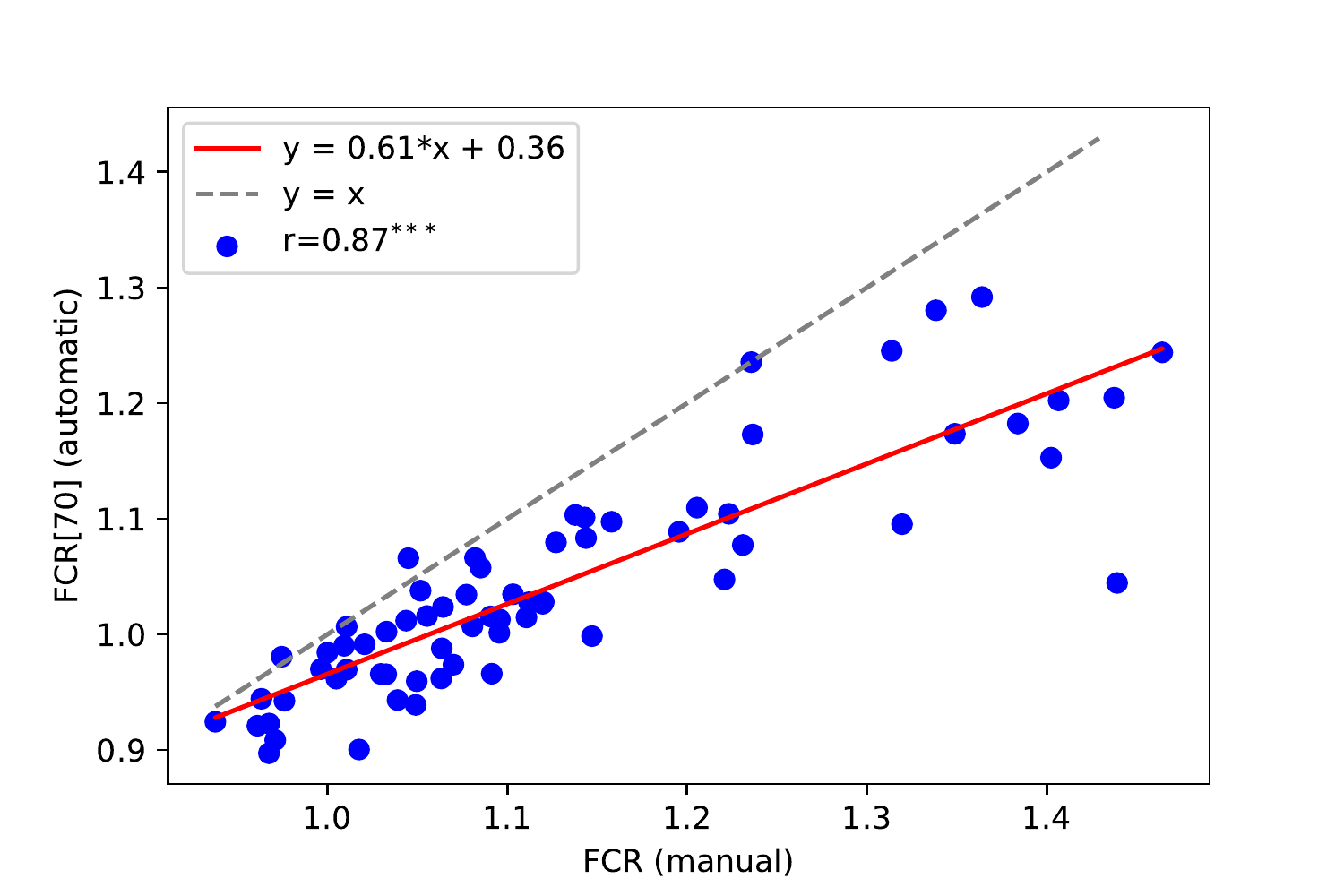}
}
\subfigure[F2i/F2u]{
\includegraphics[width=0.45\linewidth]{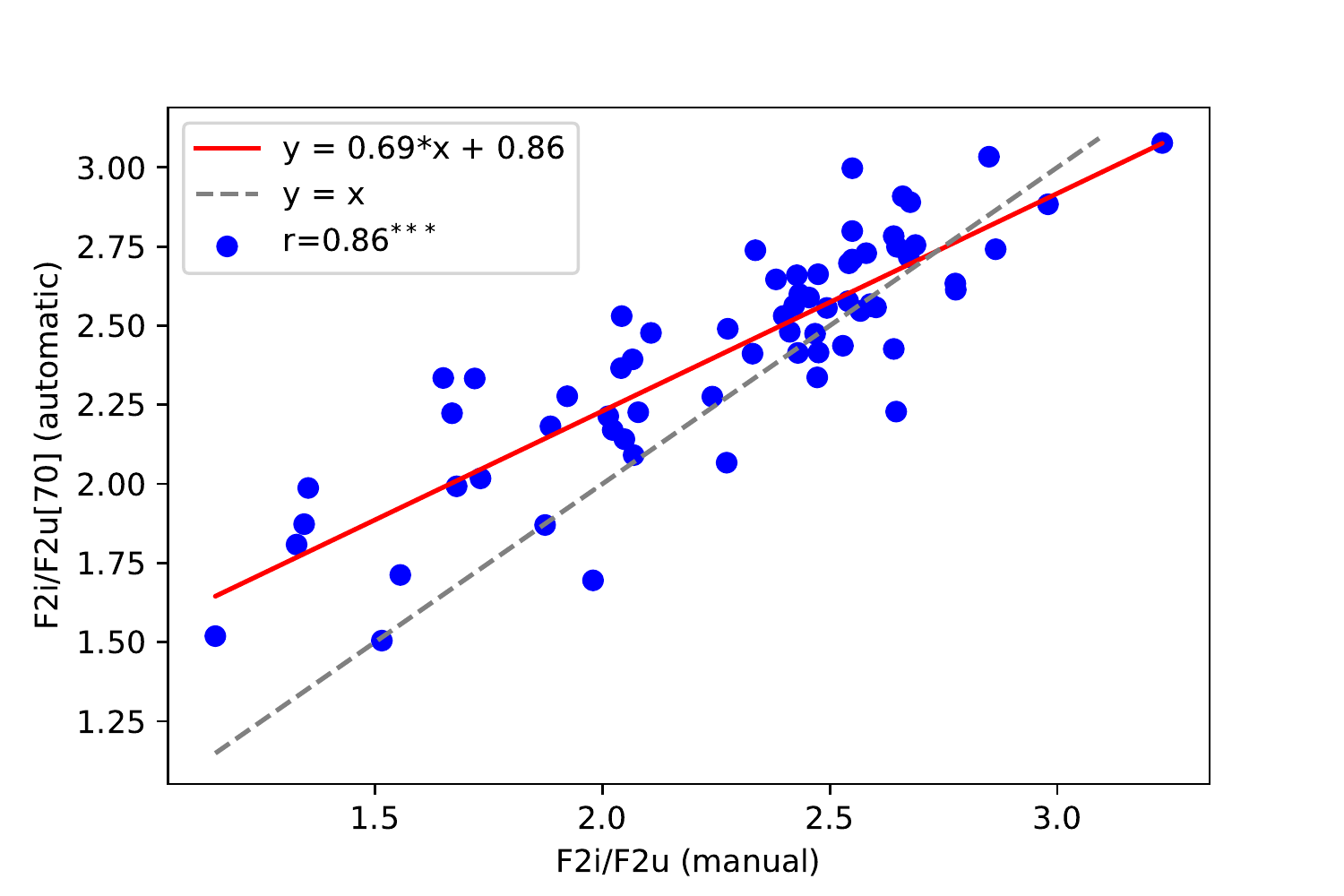}
}
\caption{Correlations of vowel articulation features computed with the mean of manually extracted formants and with 70/30 percentiles of the frame-level formants in automatic method for the 67 Finnish speakers. In the legend, `r' stands for Pearson correlation coefficient and three-stars ($^{***}$) indicates statistical significance at $p < .001$.}
\label{fig:corr_man[men]_auto[70]}
\vspace{-3mm}
\end{figure*}

Scatter plots in Fig. \ref{fig:corr_man_auto} illustrate the relationship between the automatic and manual methods on each vowel articulation, where the mean of frame-level formants are used to calculate the representative formant frequencies. In the figure, each dot (blue) represents one particular speaker. A linear regression line (red) fit to the measurements and a reference dashed line (grey) for $y=x$ are displayed together. The plots show strong correlations on features computed with automatic and manual methods. Furthermore, the dots are distributed below the dashed line ($y=x$), except those in plot of FCR. This means that the automatically computed feature values (VAI, VSA and F2i/F2u) are smaller than those from manually annotated frames. This phenomenon is reasonable, since the automatic frames selection may include phonemes which are located towards the center of the vowel space. As demonstrated by Fig. \ref{fig:corr_man[men]_auto[70]}, the centralization tendency is substantially reduced when using the percentiles (in this case 70/30) for the overall formant estimates of corner vowels in automatic selection method.

\begin{table}[!htbp]
    \centering
    \caption{Pearson correlations of formant related features between automatic and manual methods on the PDSTU corpus ($p < .00001$ for all correlations.)}
    \label{tab:corr_auto_man}
    \begin{tabular}{lllll}
        \hline
        \multirow{2}{*}{Features} & \multirow{2}{*}{Mean} & hi=50 & hi=70 & hi=90 \\
         & & lo=50 & lo=30 & lo=10 \\
        \hline
        VAI & \textbf{0.89} & 0.87 & \textbf{0.89} & 0.83 \\
        VSA & 0.87 & 0.86 & \textbf{0.88} & \textbf{0.88} \\
        FCR & \textbf{0.90} & 0.89 & 0.86 & 0.85 \\
        F2i/F2u & 0.85 & 0.83 & 0.86 & \textbf{0.87} \\
        \hline
        F1a & \textbf{0.94} & 0.93 & 0.90 & 0.91 \\
        F2a & 0.89 & 0.84 & \textbf{0.92} & 0.73 \\
        \hline
        F1i & 0.82 & \textbf{0.96} & 0.95 & 0.95 \\
        F2i & \textbf{0.97} & 0.94 & 0.95 & \textbf{0.97} \\
        \hline
        F1u & 0.85 & 0.92 & \textbf{0.94} & 0.91 \\
        F2u & 0.79 & \textbf{0.81} & 0.74 & 0.79 \\
        \hline
        Avg. (Control) & 0.86 & 0.89 & \textbf{0.90} & 0.83 \\
        Avg. (PD) & 0.85 & 0.88 & 0.87 & \textbf{0.90} \\
        Avg. (Control + PD) & 0.87 & \textbf{0.89} & \textbf{0.89} & 0.87\\
        \hline
    \end{tabular}

\end{table}

\vspace{-3mm}

\vspace{-3mm}

\subsection{Correlation between vowel articulation features and expert assessment}
As mentioned in Section \ref{sub:finnish}, experts' ratings on speech intelligibility, voice impairment and overall severity of communication disorder were collected for the $35$ PD participants and $15$ control speakers in the Finnish dataset PDSTU. Pearson correlations between vowel articulations and the average expert rating on each rated dimension were calculated and are listed in Table \ref{tab:corr_vai_vsa_assess}. 
Almost all the correlation pairs are shown to be significant.

VAI, VSA and F2i/F2u increase with the ratings of speech intelligibility but decrease with increasing voice impairment and overall severity of communication disorder. This means that articulatory undershoot increases as the speech/voice disorder severity increases. In contrast, FCR decreases with the speech intelligibility but increases with the voice impairment and overall severity of communication disorder.

These observations are compatible with previous work \cite{FCR}. The largest correlation coefficients for each pair of vowel articulation and expert rating are marked as \textbf{bold}. Absolute values of these marked coefficients range from $0.44$ to $0.68$, which are interpreted as \textbf{moderate} correlations between the vowel articulations and regarding expert rating dimension \cite{Haldun_2018}. 

The correlations obtained by automatic analyses are not different from those from manual analyses ($p > .05$; William’s test for comparing dependent correlations \cite{Williams_1959}). The only exceptions to this rule are F2i/F2u[50] and VSA[90], where there is a significant difference between manual and automated scores ($p < .05$).

Comparing the different formant estimators in terms of the resulting correlations between the articulation features and expert assessments (Table V), there are two main findings. First, the means and medians of frame-level formants generally lead to equally informative articulatory features, where only the manually computed VSA and VSA[50] show a difference for speech intelligibility (0.49 vs 0.55; $p < .05$ for William's test) and automatically computed VSA and VSA[50] have significantly different correlations with the overall severity ($-$0.56 vs $-$0.47; $p < .05$). Second, the means and medians typically lead to higher correlations than 70/30 or 90/10 percentiles across the rating dimensions. For instance, automatic VAI[50] and FCR[50] are always more correlated with the ratings than their 70/30 or 90/10 counterparts across all the rating dimensions ($p < .05$). The trend is similar for manually computed features, where means and medians perform generally better than the higher percentiles.

\begin{figure*}[htb]
    \centering
    \subfigure[VAI]{
        \includegraphics[width=0.45\linewidth]{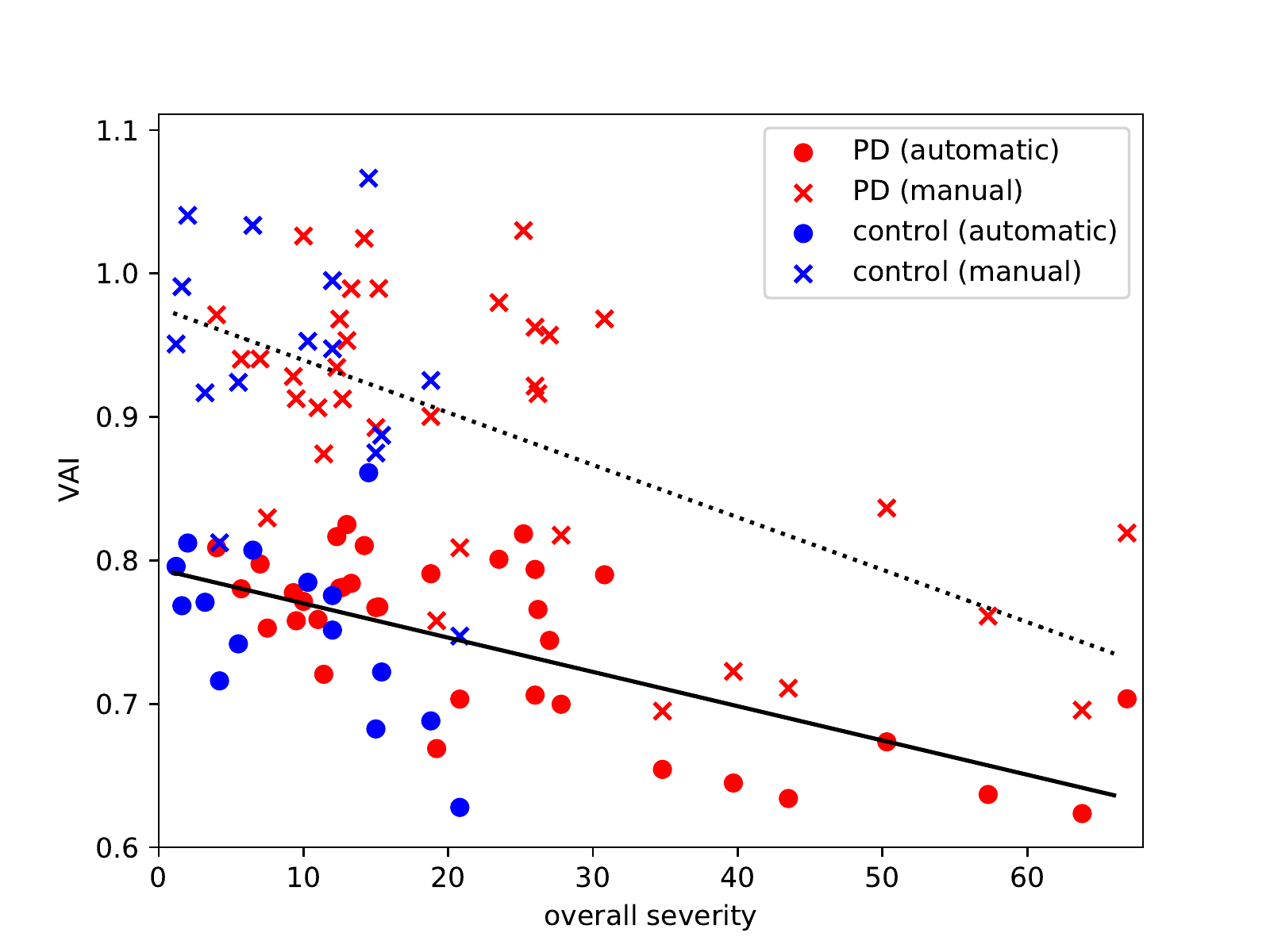}
        }
    \subfigure[FCR]{
        \includegraphics[width=0.45\linewidth]{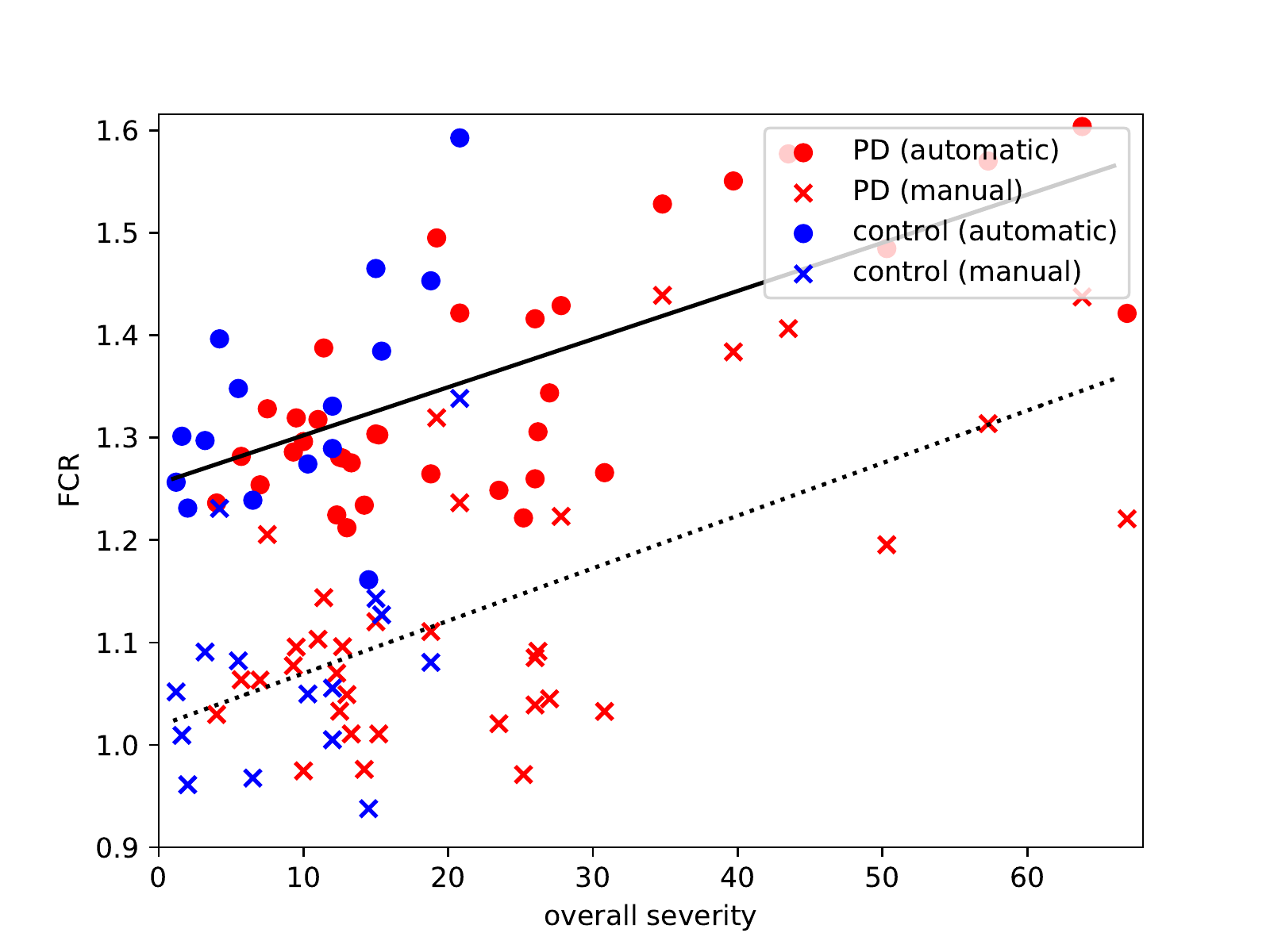}
        }
    \caption{Correlations of VAI and FCR (computed with mean of frame-level formants in automatic and manual methods) with the average of experts' ratings on overall severity of communication disorder in the Finnish database PDSTU. Here each symbol represents one speaker. Red and blue are for the PD and control groups, respectively, while dots and crosses are for automatic and manual computation methods, respectively. The solid (dotted) line shows the least-squares linear fit to the automatic (manual) measurements.}
    \label{fig:scatter_vai_speech_voice}
    \vspace{-3mm}
\end{figure*}

To better visualize the correlation between vowel articulation features and expert ratings, an example of the relationship between the automatically/manually computed VAI and FCR with average ratings of overall severity is plotted in Fig. \ref{fig:scatter_vai_speech_voice}. Negative (positive) correlation between VAI (FCR) and overall severity is clearly visible for both automatic and manual computations.

Finally, we tested whether the VAI or FCR can distinguish early PD patients with less than $2$ years from diagnosis from the healthy controls. However, VAI was not significantly lower or FCR higher for the early PDs than the controls ($p \geq$ .05; one-tailed unpaired t-test). This shows that, in contrast to \cite{speech_tasks_2013}, the sensitivity of individual articulatory space features may not always be sufficient for early-stage PD diagnostics. This is due to the natural speech articulation variability also among the healthy population, as rated by experts agnostic to the speaker's health status (see, e.g., Table II and Fig. 6).

\begin{table*}[!htbp]
    \centering
    \caption{Correlation between vowel articulation features and the experts' ratings on speech intelligibility (Intelligibility), voice impairment (Voice) and overall communication disorder (Overall) from the $35$ PD speakers and $15$ control speakers in the Finnish dataset PDSTU. One-star ($*$) denotes significance at $p < .05$, two-stars ($**$) at $p < .01$, and three-stars ($***$) at $p < .001$ (Bonferroni-corrected for 4 comparisons among the alternative formant estimation strategies).}
    \label{tab:corr_vai_vsa_assess}
    \begin{tabular}{lllllll}
        \hline
        Method & \multicolumn{3}{c}{Automatic} & \multicolumn{3}{c}{Manual} \\
        \hline
        Feature & Intelligibility & Voice & Overall & Intelligibility & Voice & Overall \\
        \hline
    VAI & $\textbf{0.60}^{***}$ & $\textbf{$-$0.57}^{***}$ & $\textbf{$-$0.63}^{***}$ & $\textbf{0.67}^{***}$ & $\textbf{$-$0.51}^{***}$ & $\textbf{$-$0.59}^{***}$ \\
    VAI[50] & $0.58^{***}$ & $\textbf{$-$0.57}^{***}$ & $-0.60^{***}$ & $0.65^{***}$ & $\textbf{$-$0.51}^{***}$ & $-0.56^{***}$ \\
    VAI[70] & $0.49^{***}$ & $-0.46^{***}$ & $-0.49^{***}$ & $0.47^{***}$ & $-0.44^{**}$ & $-0.48^{***}$ \\
    VAI[90] & $0.11^{***}$ & $-0.14^{***}$ & $-0.16^{***}$ & $0.25^{***}$ & $-0.24^{***}$ & $-0.26^{***}$ \\
    \hline
    VSA & $\textbf{0.54}^{***}$ & $\textbf{$-$0.48}^{***}$ & $\textbf{$-$0.56}^{***}$ & $0.49^{***}$ & $\textbf{$-$0.45}^{**}$ & $-0.52^{***}$ \\
    VSA[50] & $0.51^{***}$ & $-0.41^{**}$ & $-0.47^{***}$ & $\textbf{0.55}^{***}$ & $\textbf{$-$0.45}^{**}$ & $\textbf{$-$0.53}^{***}$ \\
    VSA[70] & $0.47^{***}$ & $-0.35^{*}$ & $-0.40^{**}$ & $0.43^{**}$ & $-0.35^{*}$ & $-0.44^{**}$ \\
    VSA[90] & $0.05$ & $-0.05$ & $-0.07$ & $0.12$ & $-0.12$ & $-0.20$ \\
    \hline
    FCR & $\textbf{$-$0.61}^{***}$ & $\textbf{0.58}^{***}$ & $\textbf{0.65}^{***}$ & $\textbf{$-$0.68}^{***}$ & $\textbf{0.52}^{***}$ & $\textbf{0.62}^{***}$ \\
    FCR[50] & $-0.59^{***}$ & $0.57^{***}$ & $0.62^{***}$ & $-0.65^{***}$ & $\textbf{0.52}^{***}$ & $0.58^{***}$ \\
    FCR[70] & $-0.49^{***}$ & $0.47^{***}$ & $0.50^{***}$ & $-0.46^{***}$ & $0.43^{**}$ & $0.49^{***}$ \\
    FCR[90] & $-0.15^{***}$ & $0.18^{***}$ & $0.21^{***}$ & $-0.26^{***}$ & $0.23^{***}$ & $0.27^{***}$ \\
    \hline
    F2i/F2u & $\textbf{0.56}^{***}$ & $\textbf{$-$0.57}^{***}$ & $\textbf{$-$0.59}^{***}$ & $\textbf{0.57}^{***}$ & $\textbf{$-$0.44}^{**}$ & $\textbf{$-$0.51}^{***}$ \\
    F2i/F2u[50] & $0.54^{***}$ & $-0.56^{***}$ & $-0.55^{***}$ & $0.50^{***}$ & $-0.41^{**}$ & $-0.43^{**}$ \\
    F2i/F2u[70] & $0.47^{***}$ & $-0.50^{***}$ & $-0.51^{***}$ & $0.48^{***}$ & $-0.40^{**}$ & $-0.44^{**}$ \\
    F2i/F2u[90] & $0.38^{**}$ & $-0.37^{**}$ & $-0.41^{**}$ & $0.46^{***}$ & $-0.34^{*}$ & $-0.38^{**}$ \\
    \hline
    \end{tabular}
\end{table*}

\subsection{Analysis of the Spanish PD speech}
\label{sub:exp_pcgita}
For the Spanish PC-GITA, VAI, VSA, FCR and F2i/F2u estimates were computed similarly to PDSTU using the mean and percentile estimators. For the control and PD groups, the mean and standard deviation for each feature are listed in Table \ref{tab:corr_pcgita}. Unpaired t-test was used to determine if there is a significant difference between control and PD group features, and the corresponding test outcomes and t-statistics (df = 98 for all) are reported in Table \ref{tab:corr_pcgita}.

\begin{figure}[htb]
\centering
\includegraphics[width=\linewidth]{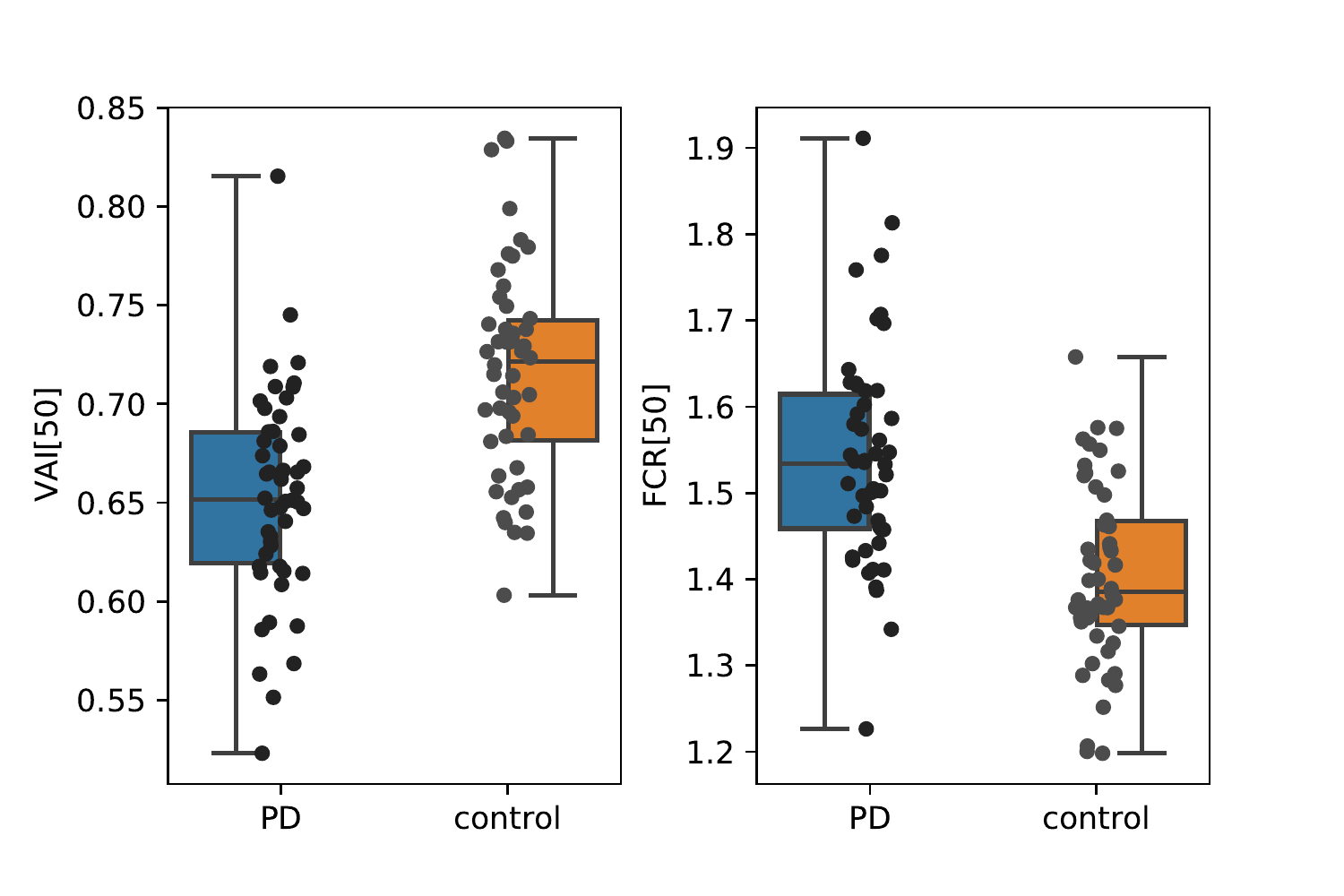}
\caption{VAI (left) and FCR (right) computed using 50/50 percentiles of frame-level formants for PD and control speakers in PC-GITA.}
\label{fig:vai_pcgita}
\vspace{-3mm}
\end{figure}

Besides comparing the feature distribution difference between control and PD groups, we also calculated the Pearson correlation coefficients for each pair of features and and ratings on UPDRS and UPDRS-speech (Table \ref{tab:corr_pcgita}, the last two columns). Here we assumed the UPDRS and UPDRS-speech to be zero for the control speakers.
The analysis shows that VAI and FCR estimates are moderately correlated with the ratings, whereas VSA turns out to be less informative. The F2i/F2u feature is correlated with the ratings, but to a less degree than VAI and FCR. Out of the different formant estimators, the median seems to be the most robust one for VAI, FCR and F2i/F2u. Fig. \ref{fig:vai_pcgita} shows the different distributions of VAI[50] and FCR[50] in PD and control groups. The obtained correlations are in a similar range to those reported in \cite{orozco2016analysis}, where VSA and FCR computed from sustained vowels /a/, /i/, and /u/ were used together with a so-called vocal pentagon area feature. When predicting the same PC-GITA UPDRS scores using a support vector regressor, they achieved a Pearson correlation of $0.41$ between the UPDRS and the predictions based on the three features.

\begin{table*}[!htbp]
    \centering
    \caption{Statistics of the automatically computed vowel articulation features for control and PD speakers in PC-GITA. In t-test of each feature between control and PD group, as well as in Pearson correlation between each feature and UPDRS or UPDRS-speech (Speech), one-star ($*$) denotes significance at $p < .05$, two-stars ($**$) at $p < .01$, and three-stars ($***$) at $p < .001$ (Bonferroni-corrected for 4 comparisons among the alternative formant estimation strategies).}
    \label{tab:corr_pcgita}
    \begin{tabular}{llllll}
    \hline
    Feature  & Control & PD & T-stat (df=$98$) & UPDRS & Speech \\
    \hline
    VAI &	$0.69\pm0.05$ &	$0.64\pm0.05$  &	$5.00^{***}$ & $-0.38^{***}$ & $-0.38^{***}$ \\
    VAI[50] &	$0.72\pm0.05$ &	$0.65\pm0.05$  &	$5.95^{***}$ & $\textbf{$-$0.46}^{***}$ & $\textbf{$-$0.43}^{***}$\\
    VAI[70] &	$0.84\pm0.06$ &	$0.77\pm0.07$  &	$5.40^{***}$ & $\textbf{$-$0.42}^{***}$ & $\textbf{$-$0.45}^{***}$\\
    VAI[90] &	$1.06\pm0.10$  &	$0.98\pm0.10$  &	$3.85^{***}$  & $-0.32^{**}$ & $-0.36^{***}$ \\
    \hline
    VSA$\times10^{-5}$ &	$2.96\pm1.60$ &	$2.50\pm2.11$ &	1.22 & $-0.07$ & $-0.10$ \\
    VSA[50]$\times10^{-5}$ &	$2.71\pm1.90$ &	$2.01\pm1.84$ &	1.87 & $-0.16$ & $-0.17$\\
    VSA[70]$\times10^{-5}$ &	$6.44\pm2.55$ &	$5.32\pm2.46$ &	$2.20^{*}$ & $-0.23^{*}$ & $-0.29^{**}$ \\
    VSA[90]$\times10^{-5}$ &	$10.32\pm3.03$ &	$10.84\pm5.84$ &	$-0.56$ & 0.07 & $-0.03$ \\
    \hline
    FCR &	$1.45\pm0.11$ &	$1.56\pm0.11$  &	$-4.97^{***}$ 	& $0.38^{***}$ & $0.39^{***}$\\
    FCR[50] &	$1.40\pm0.10$ &	$1.54\pm0.12$  &	$-5.92^{***}$ 	& $\textbf{0.45}^{***}$ & $\textbf{0.43}^{***}$\\
    FCR[70] &	$1.19\pm0.09$ &	$1.31\pm0.12$  &	$-5.51^{***}$ 	& $\textbf{0.43}^{***}$ & $\textbf{0.46}^{***}$  \\
    FCR[90] &	$0.95\pm0.09$ &	$1.03\pm0.11$  &	$-4.04^{***}$ 	& $0.33^{***}$ & $0.38^{***}$  \\
    \hline
    F2i/F2u &	$1.63 \pm 0.26$ &	$1.45 \pm 0.23$  &	$3.77^{***}$ 	& $-0.29^{**}$ & $-0.30^{**}$ \\
    F2i/F2u[50] &	$1.75 \pm 0.27$ &	$1.51 \pm 0.25$  &	$4.63^{***}$ 	& $-0.35^{***}$ & $-0.35^{***}$ \\
    F2i/F2u[70] &	$2.02 \pm 0.32$ &	$1.75 \pm 0.30$  &	$4.22^{***}$ 	& $-0.32^{**}$ & $-0.34^{***}$ \\
    F2i/F2u[90] &	$2.40 \pm 0.34$ &	$2.15 \pm 0.41$  &	$3.25^{**}$ 	& $-0.28^{**}$  & $-0.29^{**}$ \\
    \hline
    \end{tabular}
\end{table*}
\vspace{-3mm}

\subsection{Analysis of the English dysarthric speech}
\label{sub:exp_torgo}

\begin{figure*}[htb]
\centering
\subfigure[Control]{
\includegraphics[width=0.45\linewidth]{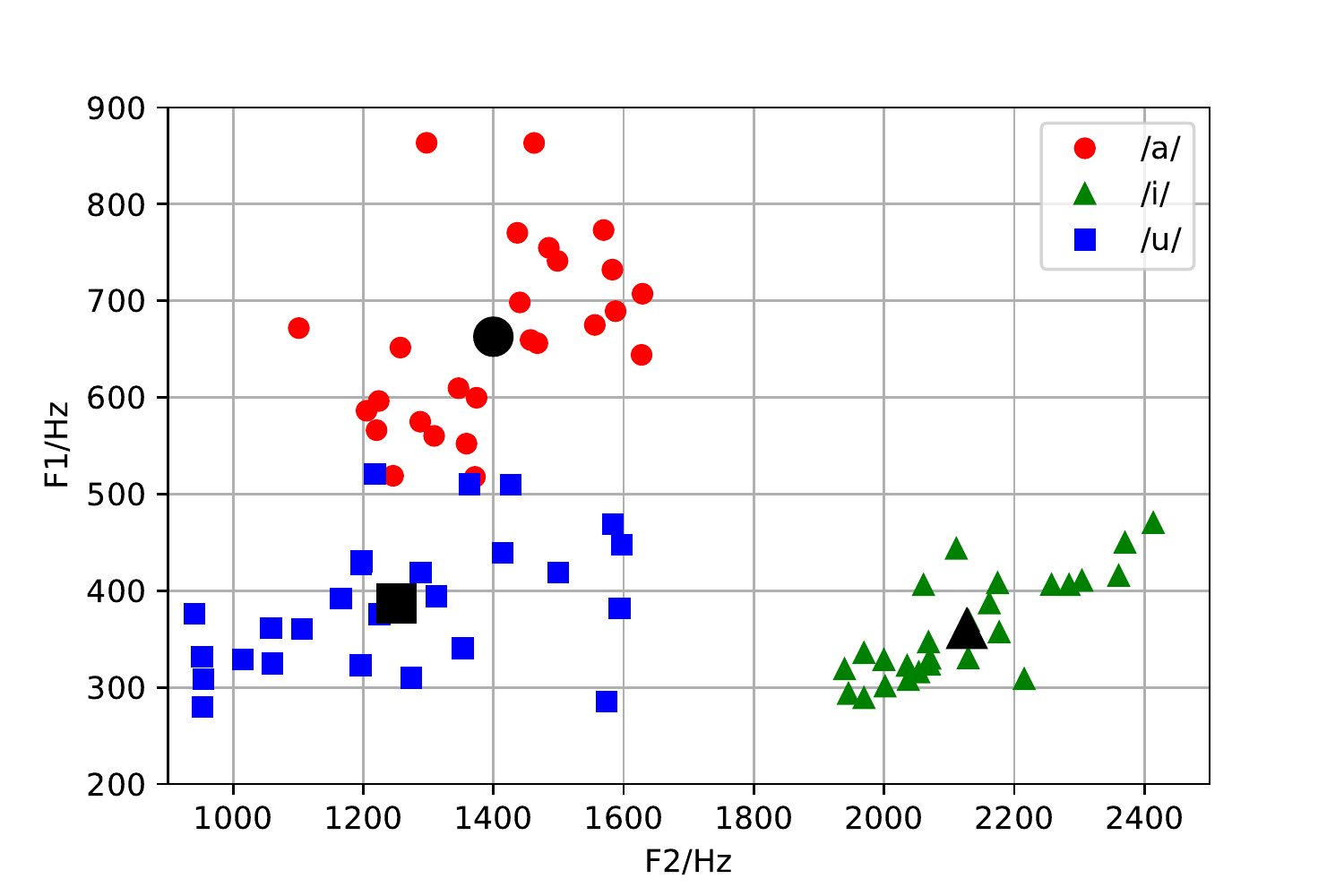}
}
\subfigure[Dysarthric]{
\includegraphics[width=0.45\linewidth]{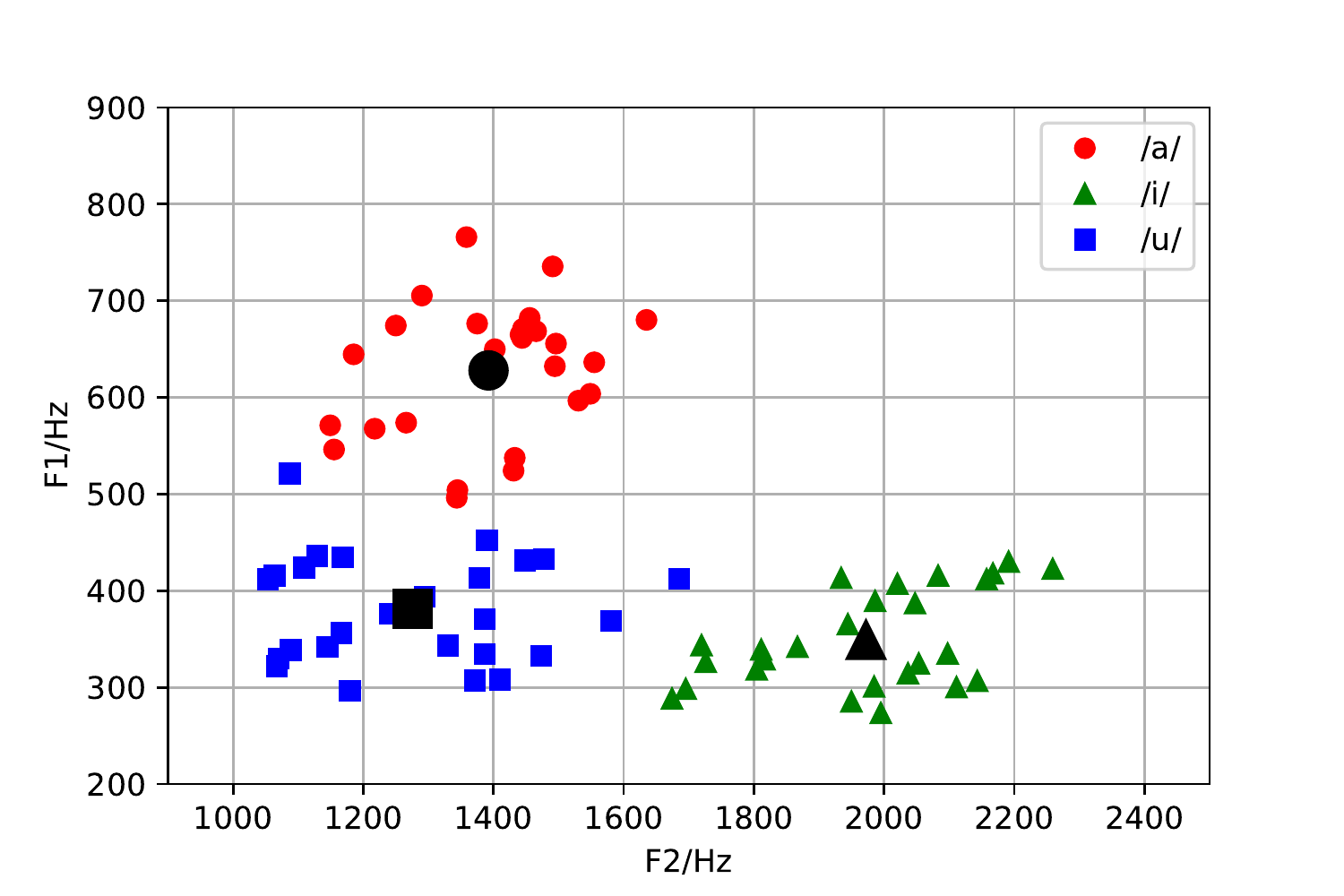}
}
\caption{Distributions of corner vowels for control and dysarthric groups in TORGO. The 70/30 percentiles of frame-level formants were used for corner vowel representation. The black symbols represent the average of each vowel category.}
\label{fig:formants_torgo}
\end{figure*}

\begin{figure}[htb]
\vspace{-6mm}
\centering
\includegraphics[width=\linewidth]{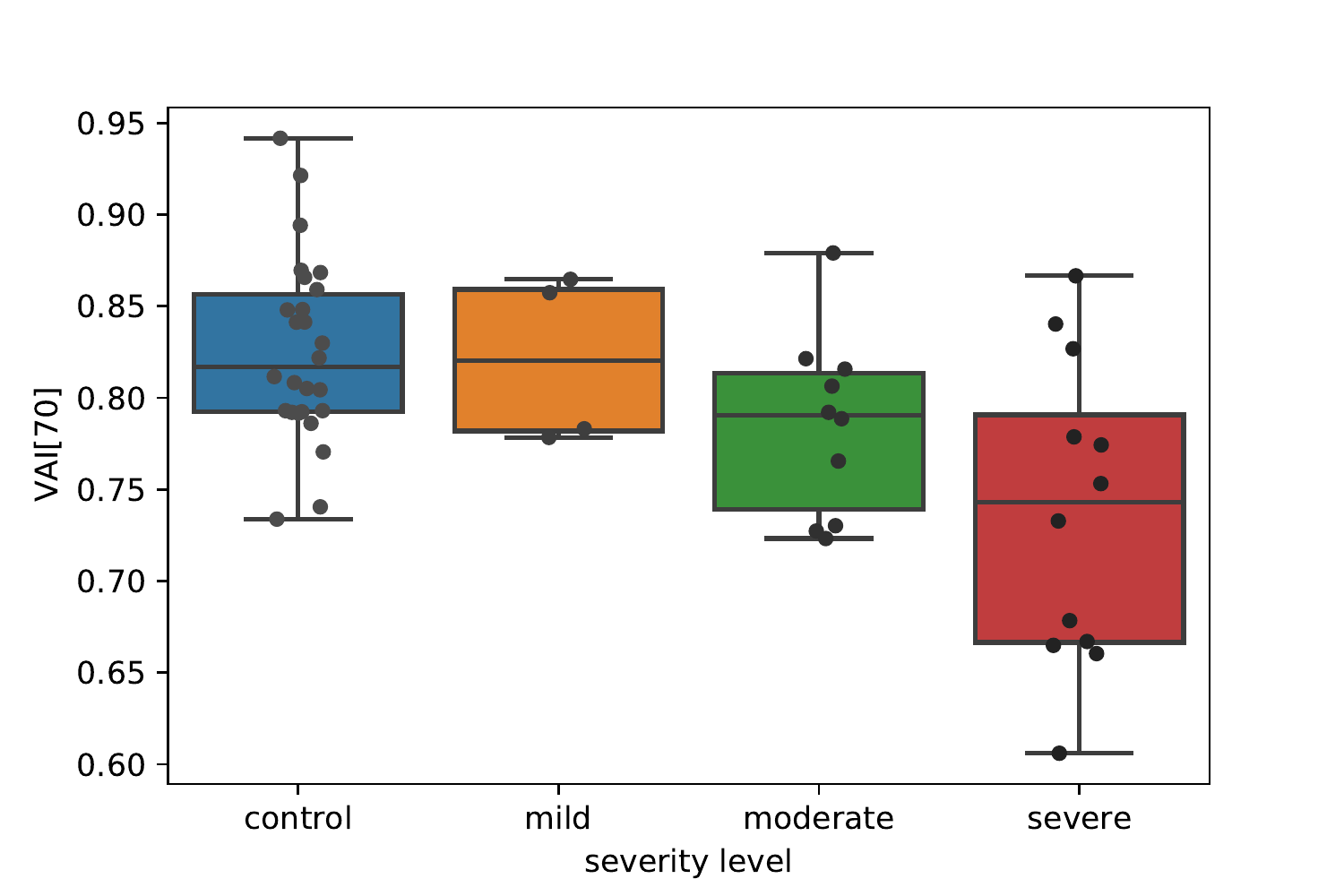}
\caption{VAI computed using 70/30 percentiles of frame-level formants for different overall severity levels in TORGO.}
\label{fig:vowel_articulations_torgo}
\vspace{-6mm}
\end{figure}

Vowel articulation parameters (VAI, VSA, FCR and F2i/F2u) were also automatically computed for each audio file in the English TORGO data, and the results were compared between the control and dysarthric groups. Fig. \ref{fig:formants_torgo} illustrates the formant estimates obtained for the two groups with the 70/30 percentile estimator. In Table \ref{tab:torgo}, the means and standard deviations are listed for the two groups on each feature. Unpaired t-test outcomes and t-statistics (df = 48 for all) are also reported for comparison of the two groups.
The last column lists the Pearson correlation coefficients between each feature and the overall dysarthria severity level. The analyses show that the differences in VAI[70], VAI[90], FCR[70], FCR[90] and F2I/F2u[70] between the two groups are significant. The correlation coefficients show that speakers with severer dysarthria tend to have smaller VAI, VSA and F2i/F2u but larger FCR values. The distribution for VAI (computed with 70/30 percentiles of formants) across different overall severity levels can also be observed in Fig. \ref{fig:vowel_articulations_torgo}. The corresponding formant estimates illustrated in Fig. \ref{fig:formants_torgo} show that this effect is primarily driven by lower average F2 of /i/ (and somewhat lower average F1 of /a/) in the dysarthric population.

\begin{table}[!htbp]
    \centering
    \caption{Statistics of automatically computed vowel articulation features for control and dysarthric speakers in TORGO. In t-test and in Pearson correlation with overall dysarthria severity level, one-star $(^{*})$, two-stars $(^{**})$ and three-stars $(^{***})$ indicate significance at $p<.05$, $p<.01$ and $p<.001$, respectively (Bonferroni-corrected for 4 comparisons among the alternative formant estimation strategies).}
    \label{tab:torgo}
    \begin{tabular}{lllll}
    \hline
     \multirow{2}{*}{Feature}  & \multirow{2}{*}{Control} & \multirow{2}{*}{Dysarthric}  & T-stat  & \multirow{2}{*}{Severity}\\
      & & & (df=48) & \\
    \hline
        VAI & $0.63 \pm 0.03$ & $0.61 \pm 0.04$ & $1.90$ & $-0.37^{**}$ \\
        VAI[50] &	$0.64 \pm 0.04$ &	$0.62 \pm 0.04$  & $1.80$ & $-0.35^{*}$\\
        VAI[70] &	$0.83 \pm 0.05$ &	$0.77 \pm 0.07$  & $\textbf{3.34}^{**}$ & $\textbf{$-$0.52}^{***}$ \\   
        VAI[90] &	$1.10 \pm 0.08$ &	$1.02 \pm 0.10$  & $\textbf{3.12}^{**}$ & $\textbf{$-$0.47}^{***}$ \\
    \hline
        VSA$\times10^{-5}$ & $3.99 \pm 2.18$ & $3.45 \pm 1.51$ & 1.01 & $-0.23$ \\
        VSA[50]$\times10^{-5}$ &	$3.57 \pm 2.03$ &	$3.63 \pm	1.95$  &	$-0.10$  & $-0.11$\\        
        VSA[70]$\times10^{-5}$ &	$8.59\pm3.06$ &	$7.51 \pm	1.97$  &	1.48 & $-0.29^{*}$ \\
        VSA[90]$\times10^{-5}$ &	$14.43\pm 4.93$ &	$12.05\pm	3.81$  &	1.91 & $-0.29^{*}$\\
    \hline
        FCR & $1.59 \pm 0.07$ & $1.64 \pm 0.10$ & $-1.96$ & $0.38^{**}$ \\
        FCR[50] &	$1.57 \pm 0.09$ &	$1.62\pm 0.11$ &	$-1.84$ & $0.36^{**}$\\
        FCR[70] &	$1.22 \pm 0.07$ &	$1.31\pm 0.13$  &	$\textbf{$-$3.34}^{**}$ & $\textbf{0.54}^{***}$\\        
        FCR[90] &	$0.91 \pm 0.07$ &	$0.99 \pm 0.10$ &	$\textbf{$-$3.15}^{**}$ & $\textbf{0.49}^{***}$\\
    \hline
        F2i/F2u & $1.29 \pm 0.11$ & $1.24 \pm 0.13$ & $1.34$ & $-0.26$ \\
        F2i/F2u[50] &	$1.34 \pm 0.18$ &	$1.26 \pm 0.16$  &	$1.74$ & $-0.28^{*}$\\    
        F2i/F2u[70] &	$1.74 \pm 0.24$ &	$1.57 \pm 0.24$  &	$\textbf{2.43}^{*}$ & $-0.36^{**}$\\
        F2i/F2u[90] &	$2.23 \pm 0.29$ &	$2.10 \pm 0.36$  &	$1.33$ & $-0.19$\\
    \hline
    \end{tabular}

\end{table}
\vspace{-3mm}

\section{Error analysis}
\label{sec:discuss}

The experimental results in Section \ref{sec:result} confirm that the proposed approach achieves comparable accuracy to manual procedure on the Finnish dataset, and that the approach is also applicable to Spanish PD speech as well as English dysarthric speech analysis without adjusting the system parameters.

However, in order to better understand the behaviour of the automatic pipeline, this section describes more detailed error analyses at different stages of the processing pipeline for the Finnish PD speech corpus.
\subsection{Efficacy of automatic candidate frames selection}
\label{sub:disc_frames_selection}
Computation of vowel articulation features relies on the selection of speech frames related to corner vowels. In the manual method, ten segments were extracted from fixed words for each corner vowel, even though the actual occurrence counts of the corner vowels are higher in the reading material. On the other hand, the automatic method attempts to find all the speech frames having similar acoustic characteristics with corner vowels. Obviously, if more corner-vowel-related phonemes occur in the reading material, the automatic method tends to select more speech frames. Moreover, variations in speaking rate and speech/voice impairment severity among speakers can have an impact on automatic frames selection.

Statistics of the automatically and manually selected frame counts and vowel occurrences in the reading material for the 67 Finnish speakers are shown in Table \ref{tab:stat_candidate_frames}. As expected, the automatic method selects more speech frames than manual annotation. The automatically selected frame counts for /a/ and /i/ are much larger than that for /u/. This can be mainly attributed to the differences in phoneme occurrence counts (introduced in Section \ref{sub:finnish}). However, /a/ has more frames selected than /i/, even though it is less frequent in terms of number of tokens in the passage. It should be noted that there are more `close' phonemes included in $\textbf{Z}_a$ than $\textbf{Z}_i$ and $\textbf{Z}_u$, as shown in Table \ref{tab:phone_sets}. In addition, phoneme /AE/ (written as `ä' in Finnish) is also assigned to $\textbf{Z}_a$, and it occurs $19$ times in the reading passage. These explain that vowel /a/ gets the highest number of automatically selected frames.

We also compared the amount of detected corner vowel frames with speaking rate and speech/voice disorder ratings using Pearson correlation. The analysis revealed that the selected frame counts for /a/ and /i/ were moderately correlated with duration of the samples (i.e., negatively correlated with the average speaking rate) with $r=0.38$ and $p<.01$. In contrast, the frame counts were not correlated with the dimensions of expert assessment on speech and voice ($p > .05$ for all comparisons), i.e., the frame selection was not affected by the severity of dysarthria in our patient group.

\begin{table}[!htbp]
    \centering
    \caption{Counts of the selected corner-vowel-related frames and corresponding phoneme occurrence counts in the Finnish reading material (the mean $\pm$ one standard deviation). `Automatic$^{\dag}$' corresponds to use only 3 corner vowels for automatic candidate frames selection, rather than extended phone categories.} One frame corresponds to approximately 30-ms of speech.
    \label{tab:stat_candidate_frames}
    \begin{tabular}{llll}
        \hline
         & /a/ & /i/ & /u/ \\
        \hline
        Automatic & $136\pm13$ & $101\pm12$ & $38\pm9$ \\
        Automatic$^{\dag}$ & $20\pm5$ & $56\pm9$ & $17\pm7$ \\
        % & 13 & 14 & 9.4 \\
        % \hline
        Manual & $28\pm2$ & $27\pm2$ & $28\pm2$ \\
        % & 2 & 2 & 2 \\
        % \hline
        Occurrences & 46 & 61 & 22 \\
        \hline
    \end{tabular}
\end{table}

Besides the selected frame counts, we computed the percentage of manually annotated frames which are also automatically selected for the same vowel. Approximately $33\%$ of manually annotated frames for /a/ and /u/ were covered by the automatic selection. For /i/, the coverage increases to $54\%$. In terms of the annotated segments (10 per vowel), more than $85\%$ of the annotated segments at least partially overlap with the automatically selected frames. The number of overlapping segments was $9.4\pm 1$, $9.1\pm 1$ and $7.1\pm 2$ for /a/, /i/ and /u/ among all the $67$ Finnish participants. Analysis on Table \ref{tab:stat_candidate_frames} also shows that the number of overlapping segments of /u/ varies widely across different speakers. For example, less than half of the annotated segments of /u/ have partial overlap with the automatic ones for two speakers. This heightened variation is generally in line with the small average and high variance in the detected /u/ frame counts.

\subsection{Benefits of using the extended phone categories}
\label{sub:disc_wide_phones}

In our proposed method for automatic corner vowel frames selection described in Section \ref{sub:screen}, each corner vowel was defined in terms of an extended set of phone recognition output categories in order to account for automatic phone recognition errors in normal and especially in dysarthric speech.
In order to investigate the benefits of the extended categories, we compared extended categories to an automatic system where only the three corner vowel recognition outputs /a/, /i/ and /u/ were used as a basis for formant estimation.

\begin{figure}
    \centering
    \includegraphics[width=\linewidth]{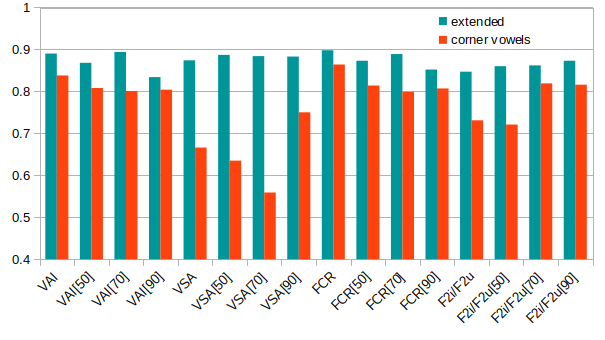}
    \caption{Pearson correlation coefficients between the manually and automatically computed vowel articulation features with the extended phone recognition categories (``extended'') and with only considering the three corner vowel output categories in the formant estimation process (``corner vowels''). }
    \label{fig:extend_small}
    \vspace{-3mm}
\end{figure}

As shown in Table \ref{tab:stat_candidate_frames}, much fewer frames were selected for each corner vowel than using the extended phone categories. The corresponding automatically computed vowel articulation features were also found to be less correlated with the ones computed with manual annotations than the features extracted using extended phone sets (Fig. \ref{fig:extend_small}). As can be seen from the figure, the drop in correlation is substantial for some of the features, such as VSA or F2i/F2u. This unanimously demonstrates the benefit of using extended phone sets in the analysis of phone recognition output.

\vspace{-3mm}
\subsection{Corner vowel formant distributions}
\begin{figure*}[htb]
\centering
\subfigure[mean formants (manual)]{
\includegraphics[width=0.45\linewidth]{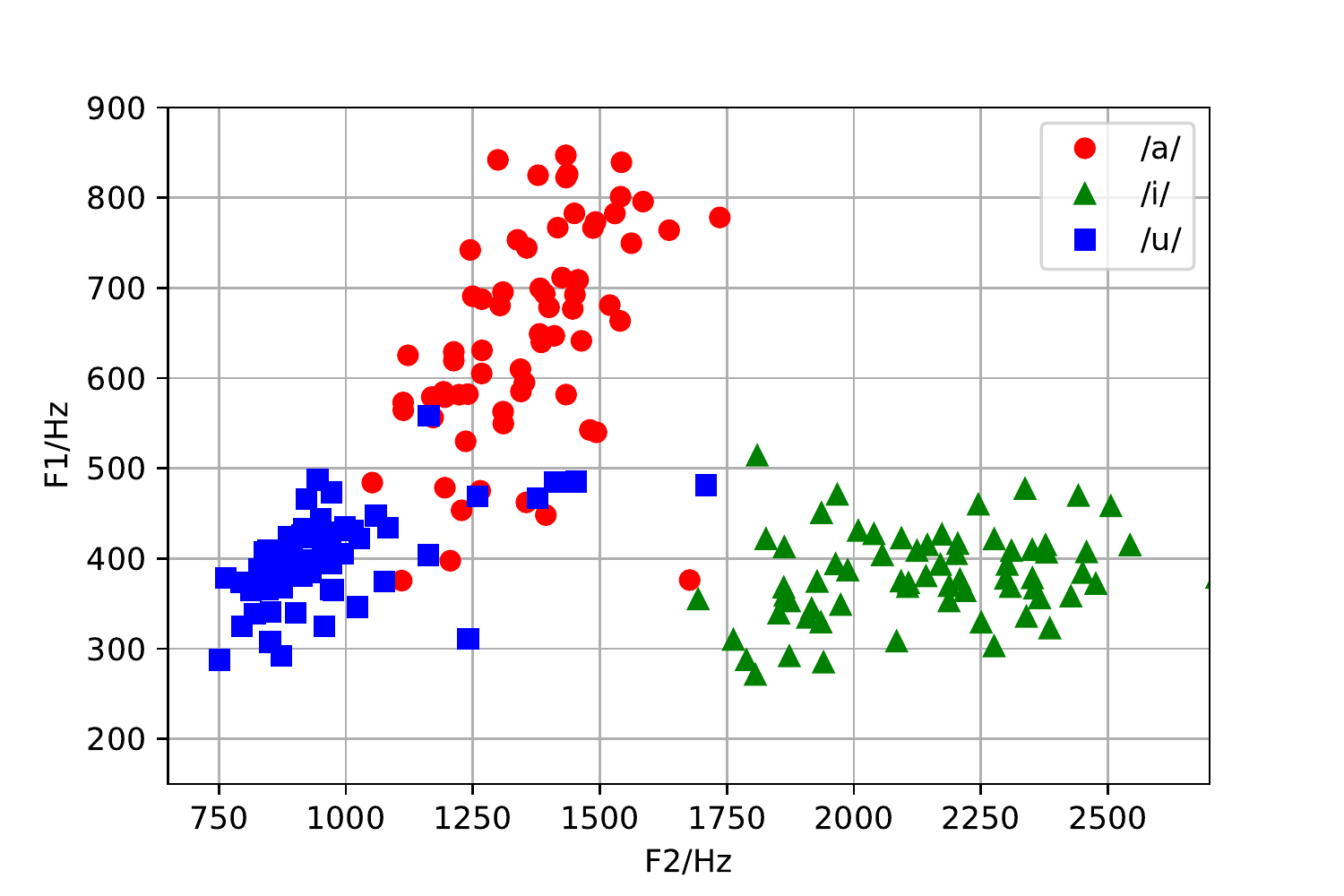}
\label{fig:mean_man}
}
\vspace{-1mm}
\subfigure[mean formants (automatic)]{
\includegraphics[width=0.45\linewidth]{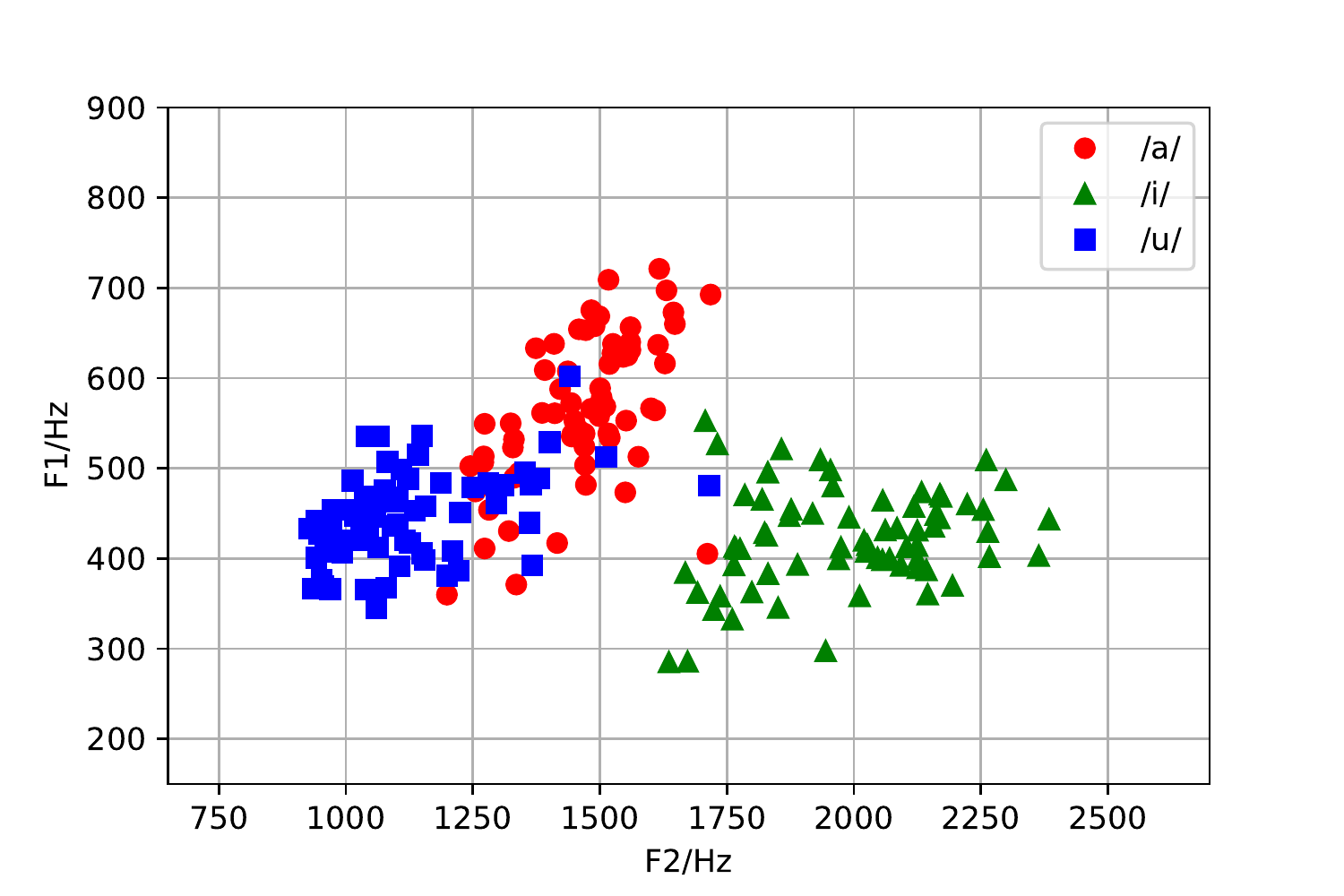}
\label{fig:auto_mean}
}
\vspace{-1mm}
\quad
\subfigure[50/50 percentiles of formants (automatic)]{
\includegraphics[width=0.45\linewidth]{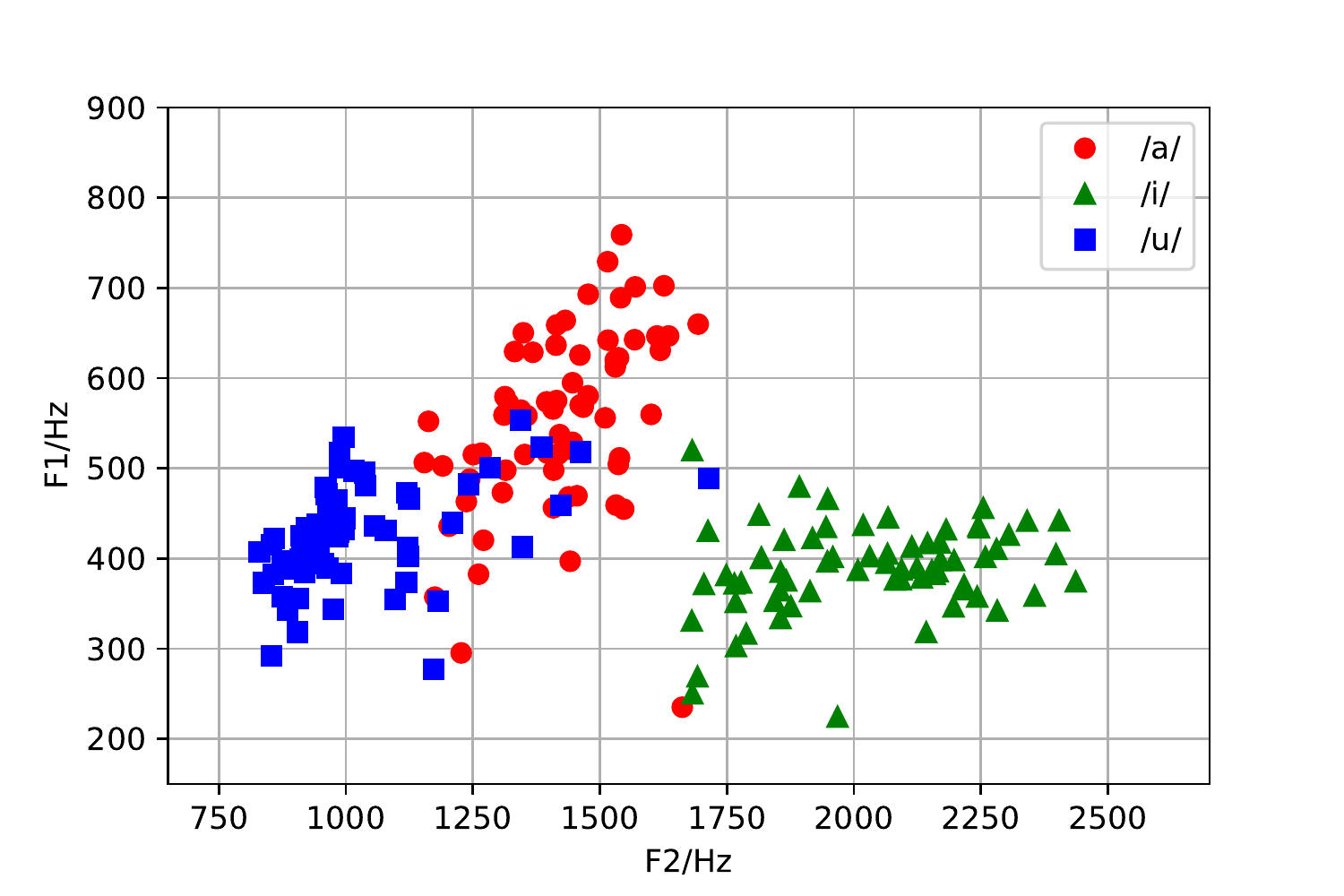}
\label{fig:auto_50}
}
\subfigure[70/30 percentiles of formants (automatic)]{
\includegraphics[width=0.45\linewidth]{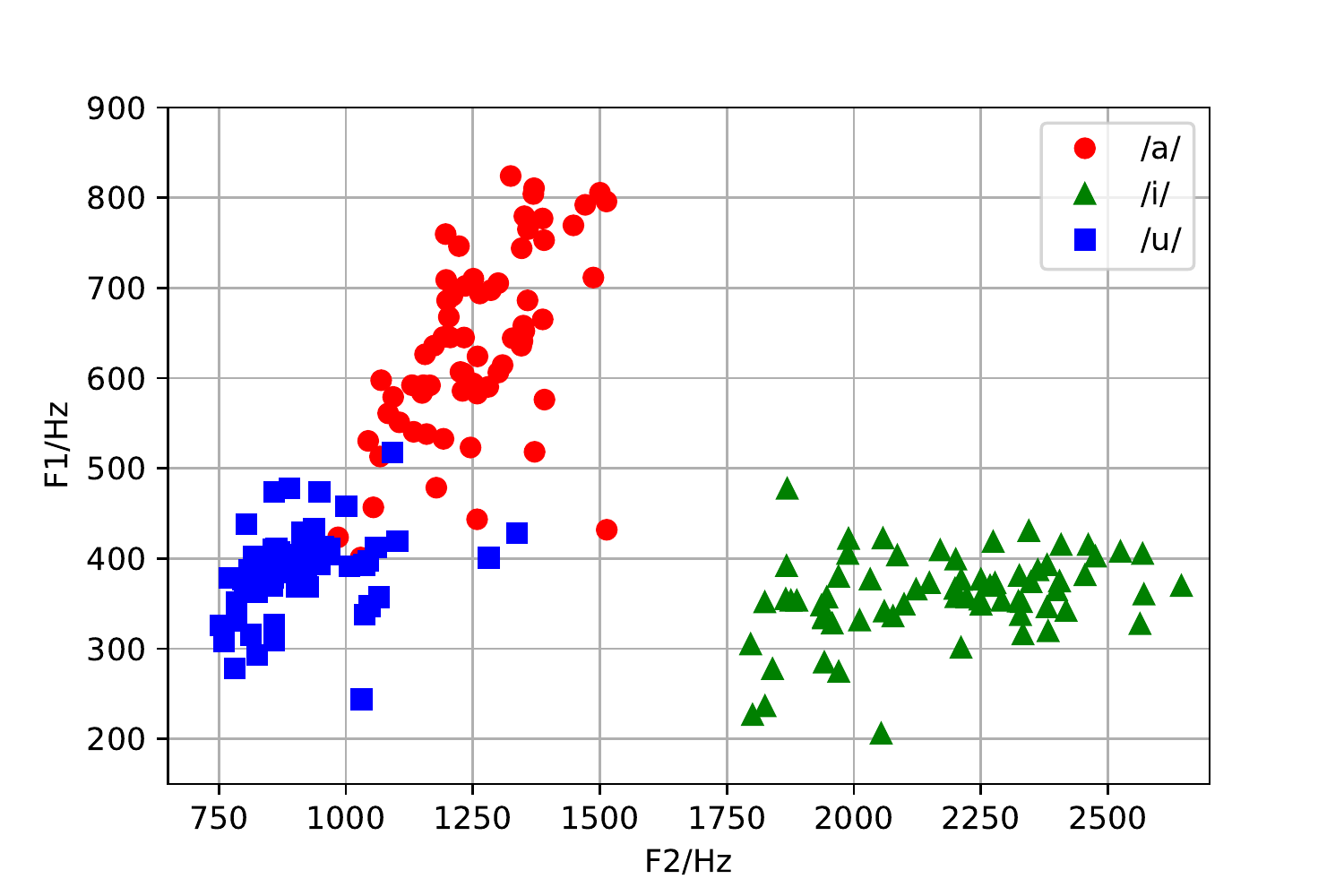}
\label{fig:auto_70}
}
\caption{Distributions of corner vowels for the $67$ Finnish participants in PDSTU.}
\label{fig:formants_dist}
\end{figure*}

\begin{figure}[htb]
    \centering
    \includegraphics[width=\linewidth]{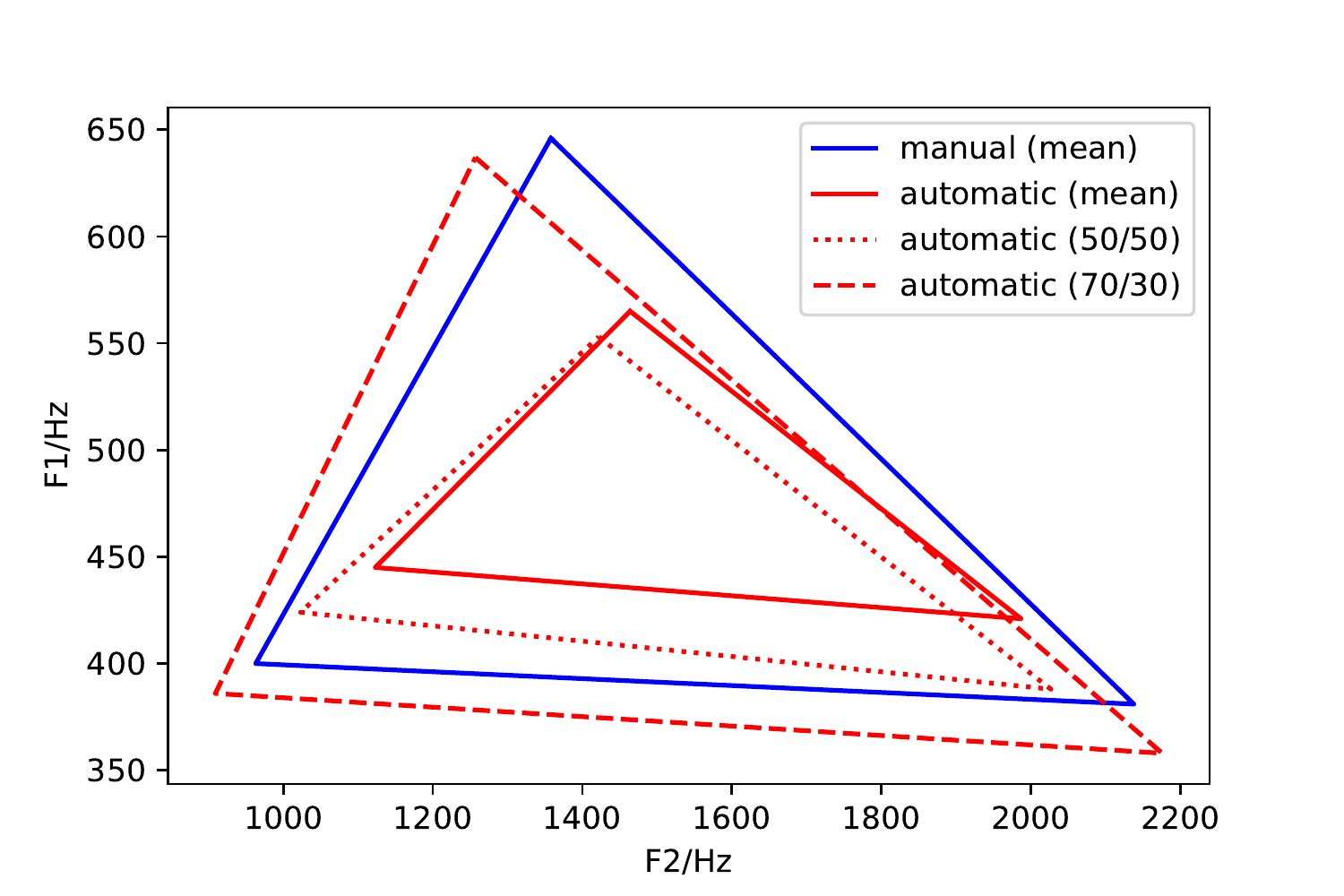}
    \caption{Average vowel articulation space of all $67$ Finnish participants in PDSTU. For manual annotation, corner vowels of each speaker were represented by mean formants. For automatic method, corner vowels of each speaker were represented by mean or 50/50 or 70/30  percentiles of frame-level formants.}
    \label{fig:corner_vowels}
    \vspace{-3mm}
\end{figure}

In addition to the reliability of frames selection, vowel articulation feature computation can be affected by formant estimates. 
Fig. \ref{fig:formants_dist} shows the average locations of corner vowels for all the $67$ Finnish speakers in the F1-F2 space. In all plots, each symbol (dot, square or triangle) represents a corner vowel formant estimate of a single speaker. For manual method, a corner vowel is represented by the mean of F1 and F2 estimates of annotated segments, as shown in Fig. \ref{fig:mean_man}. For the automatic method, corner vowels represented by the mean, the 50/50 and 70/30 percentiles of formants of selected frames are depicted in Fig. \ref{fig:auto_mean}, \ref{fig:auto_50} and \ref{fig:auto_70}. The figure shows that the three corner vowels are well separated with both manual annotation and automatic selection. For automatic selection, the distance between /a/ and /u/ can be enlarged by representing the corner vowels with apices of frame-level F1 and F2. Average positions of each corner vowel among all participants under the four conditions in Fig. \ref{fig:formants_dist} are illustrated as vertexes of triangles in Fig. \ref{fig:corner_vowels}. The figure shows that vowel articulation space represented by 70/30 percentiles of frame-level formants (red dashed lines) is the most similar to the manual one (solid blue lines). Also, the centralization tendency of the mean frame-level estimate in automatic selection is clearly observable in the figure (red solid lines).

In this work, frame-level formants were computed from  linear predictive coefficients which were calculated with Burg's algorithm \cite{burg} from speech signal by Praat. When the formants are at low frequencies or close to each other, for example in case of low back vowel /a/, formant estimation is prone to errors \cite{formant_error}. In order to estimate the number of gross estimation errors, a straightforward criterion was designed based on acoustically and articulatorily feasible formant values. More specifically, we first determined the highest feasible F1 and F2 values for /a/ vowel across a population of speakers in \cite{english_vowels}, i.e. extreme exemplar of /a/ (1002 Hz, 1688 Hz). The F1 and F2 values of extreme exemplar /a/ were set as boundaries and frames with concurrent F1 and F2 beyond the predefined boundaries were counted as frames with formant estimation errors.

\begin{figure*}[htb]
    \centering
    \subfigure[automatic selection]{
    \includegraphics[width=0.45\linewidth]{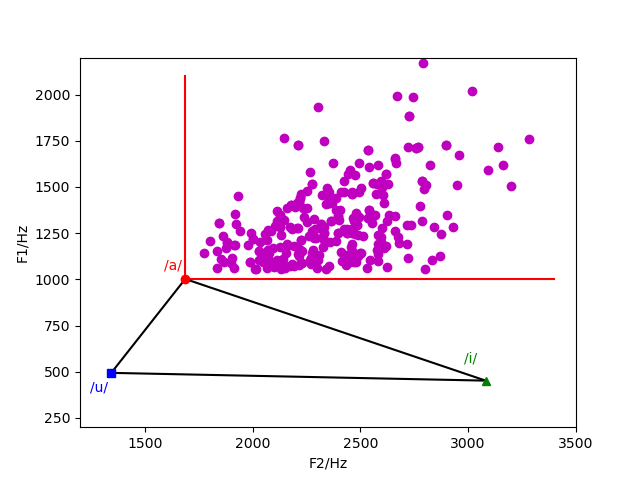}
    }
    \subfigure[manual annotation]{
    \includegraphics[width=0.45\linewidth]{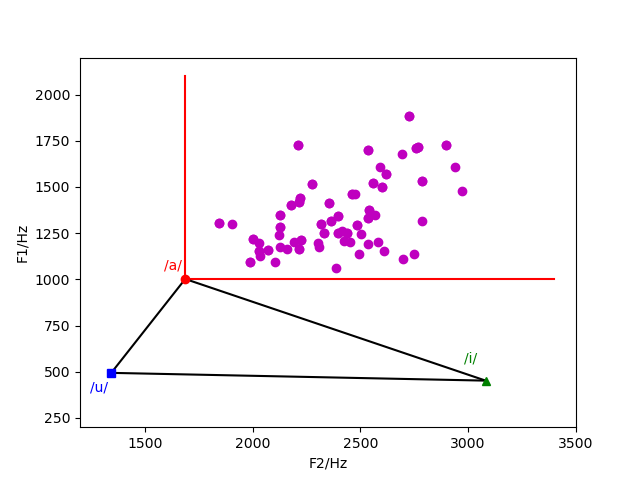}
    }
    \caption{Selected frames with formant estimation errors for all $67$ Finnish participants. Each purple dot represents one single frame. Extreme exemplars of corner vowels /a/, /i/ and /u/, as determined based on \cite{english_vowels}, are shown as a reference.}

    \label{fig:formant_errors}
    \vspace{-3mm}
\end{figure*}

In Fig. \ref{fig:formant_errors}, the exemplars of three corner vowels are marked together with the red lines indicating the error detection boundaries. In the error area, frames tend to have large absolute frequencies and small mutual differences between F1 and F2. Frames with formant estimation errors are depicted as purple dots for automatic selection (left) and manual annotation (right). Among all the Finnish participants, an 82-year old male speaker got the highest ratio of formant estimation errors among his selected frames, corresponding to $6\%$ of erroneous frames for automatic selection and $8\%$ for manual annotation. It seems that the ratio of formant estimation errors is low such that it makes no significant difference on experimental results after removing the error frames.

\section{Conclusion and future work}
\label{sec:conclude}
The present study proposed a language-independent, reliable and efficient method of automatic vowel articulatory undershoot quantification from corner vowels with the help of a universal phone/phoneme recognizer. Such an approach is highly useful for clinical work with dysarthric patients, where quantitative analysis of articulation capability typically involves laborious hand-annotation of vowel segments of interest. 

The efficacy and reliability of our proposed automatic computation of vowel articulation parameters was tested and verified on three speech corpora in different languages. We demonstrated on a Finnish PD corpus that vowel articulation features computed with automatic speech frame selection have strong correlations with the same features computed using manual annotations. In terms of correlations with subjective expert assessments of speech intelligibility, voice impairment and overall severity of communication disorder, the automatically computed features had comparable correlations as with those computed manually. However, VAI and FCR alone were not able to discriminate early-stage PD patients from controls in PDSTU. On the Spanish PD corpus PC-GITA and English dysarthria corpus TORGO, the distributions of automatically computed vowel articulation features were significantly different between the control and PD/dysarthric groups. In addition, automatically computed vowel articulation features VAI and FCR were moderately correlated with ratings of UPDRS and UPDRS-speech in PC-GITA and with overall dysarthria severity levels in TORGO.
For two of the corpora (PC-GITA and TORGO), the results also show a fair advantage of using median or percentiles 70/30 of frame-level formants over using mean formants. 
Finally, we performed error analyses of our system and showed how the use of extended phone sets to decode the automatic recognizer output improves robustness of the acoustic feature estimates. 

In the future, we plan to validate the proposed method on spontaneous speech. In addition, our goal is to explore new parametrizations of speech that would simultaneously be transparent to clinicians while having maximal sensitivity to early signs of neurodegenerative diseases in speech.

\section{Acknowledgment}
OR was funded by Academy of Finland grant no. 314602.

\normalem
\bibliographystyle{IEEEtran}
\bibliography{yuanliu}

% \iffalse

\begin{IEEEbiography}[{\includegraphics[width=0.8in,height=1.0in,clip,keepaspectratio]{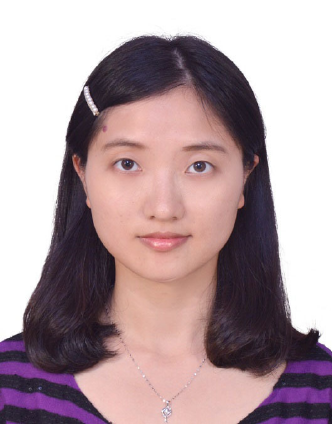}}]{Yuanyuan Liu} was born in China in 1988. She received B.S. in techniques and science of electronics in 2010 from the Central South University in Changsha, China, and Ph.D. degree in electronic engineering from the Chinese University of Hong Kong, Hong Kong, in 2019.

From 2010-2014, she was a full-time senior engineer in department of wireless products in MediaTek Inc.(Shenzhen). From 2014-2018, she was a part-time research assistant in the Shenzhen Municipal Engineering Laboratory of Speech Rehabilitation Technology in China. She is now a postdoctoral researcher at the Unit of Computing Sciences at Tampere University, Finland. Her current research includes pathological speech signal analysis and automatic assessment, automatic speech recognition and machine learning.
\end{IEEEbiography}
\vspace{-6mm}

\begin{IEEEbiography}[{\includegraphics[width=0.8in,height=1.0in,clip,keepaspectratio]{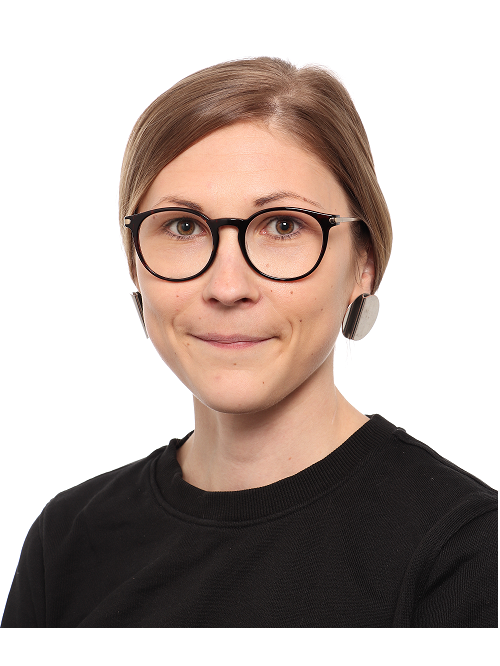}}]
{Nelly Penttilä} was born in Finland in 1989. She received the M.Sc. degree in logopedics from Tampere University, Finland, in 2012, and Ph.D. degree in logopedics from the Tampere University, in 2019. She has received multiple awards from different foundations, e.g. Fulbright Finland Foundation, and Emil Aaltonen Foundation. She currently works as a senior lecturer in the discipline of logopedics and as a principal investigator in Kuuluva Ääni -project (carrying voice-project). Her research interests include fluency, fluency disorders and speech intelligibility. She also works as speech and language pathologist.
\end{IEEEbiography}
\vspace{-6mm}
\begin{IEEEbiography}[{\includegraphics[width=0.8in,height=1.0in,clip,keepaspectratio]{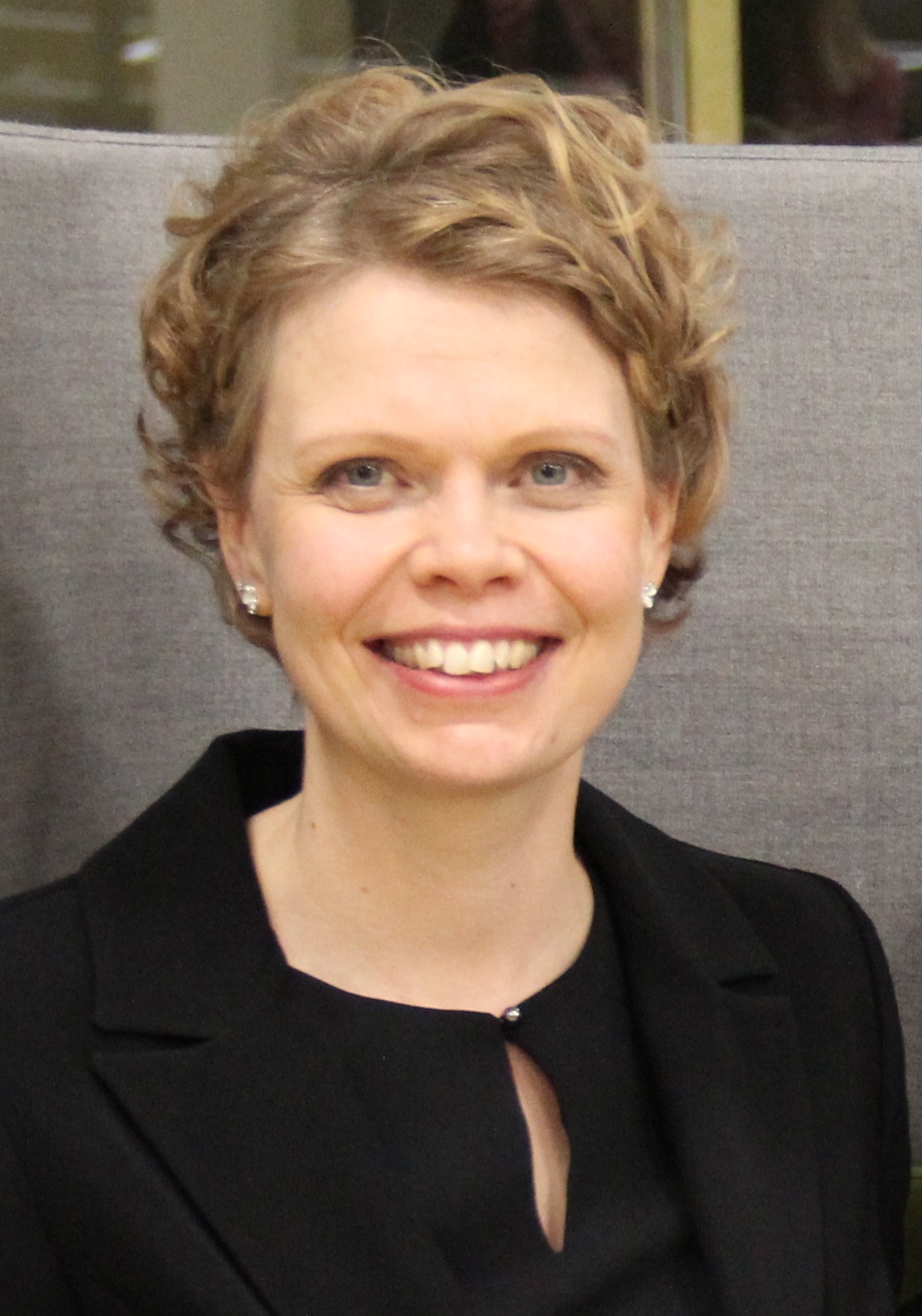}}]
{Tiina Ihalainen} was born in Finland in 1974. She received the M.Sc. degree in logopedics from Oulu University, Finland, in 2008, and Ph.D. degree in logopedics from the Tampere University, Finland, in 2018. From 2008 to 2019, she worked as a speech therapist at the Tampere University Hospital, Finland. She is specially experienced in working with adult people who have neurological diseases affecting in speech and communication (e.g. stroke, brain injury, brain tumor, Parkinson’s disease, motor neuron disease, different types of dementia and other neurogenerative diseases.). Currently, she is a senior lecturer in Degree Program in Logopedics at the Faculty of Social Sciences in the Tampere University. Her current research includes perceptual and automatic assessment of typical and pathological speech and creating novel user interfaces for augmentative and alternative communication (AAC). 
\end{IEEEbiography}
\vspace{-6mm}
\begin{IEEEbiography}[{\includegraphics[width=0.8in,height=1.0in,clip,keepaspectratio]{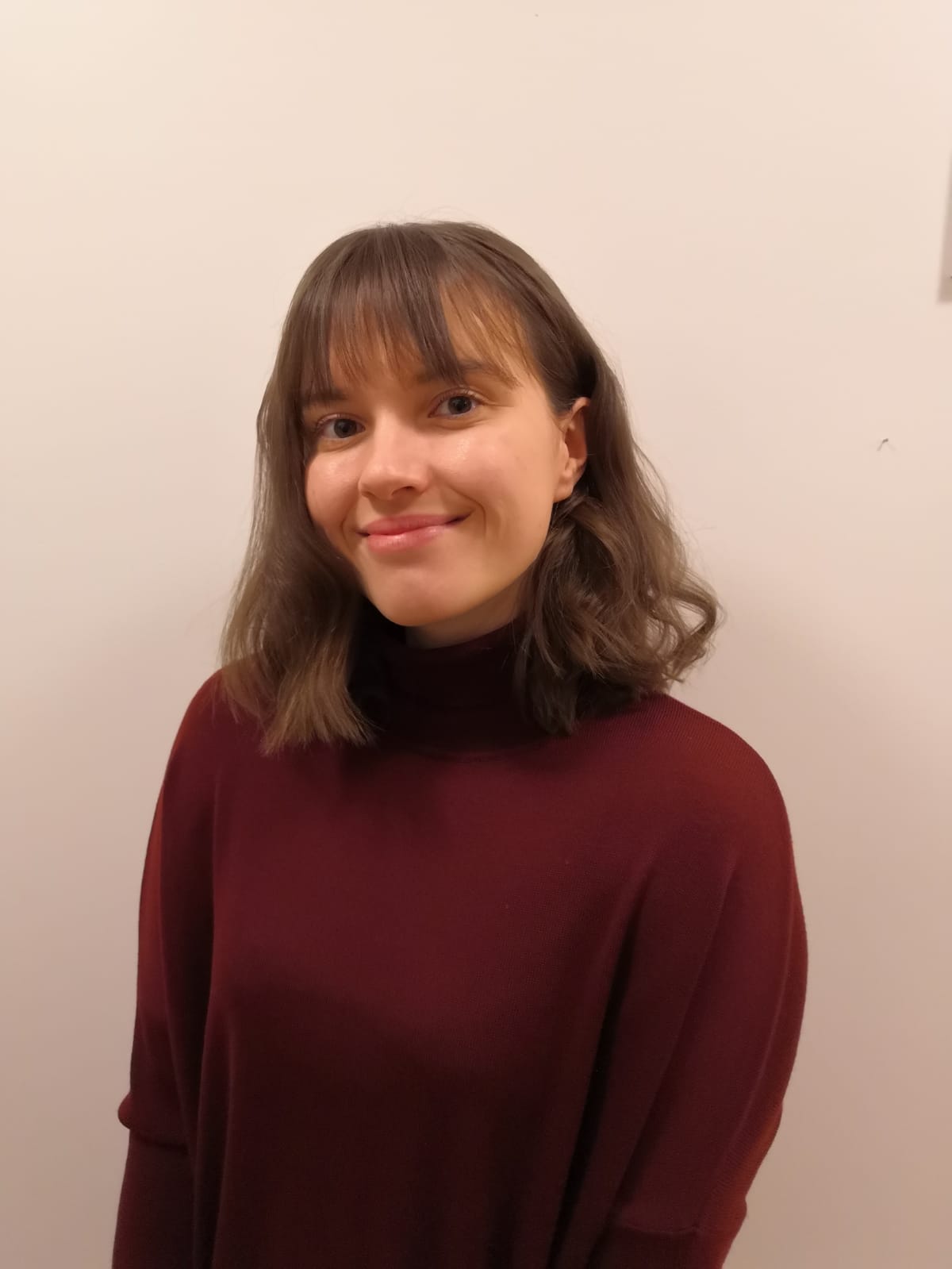}}]
{Juulia Lintula} was born in Finland in 1995. She received a Bachelor’s degree in logopedics from Tampere University, Finland, in 2019, and is now finishing her studies in the Master’s program in logopedics. In her Bachelor’s and Master’s theses she focused on speech intelligibility in Parkinson’s disease and used the vowel articulation index as a measure. She worked as a research assistant in Tampere University and collected speech data from healthy adult speakers. She currently works as a speech and language pathologist intern.
\end{IEEEbiography}
\vspace{-6mm}
\begin{IEEEbiography}[{\includegraphics[width=0.8in,height=1.0in,clip,keepaspectratio]{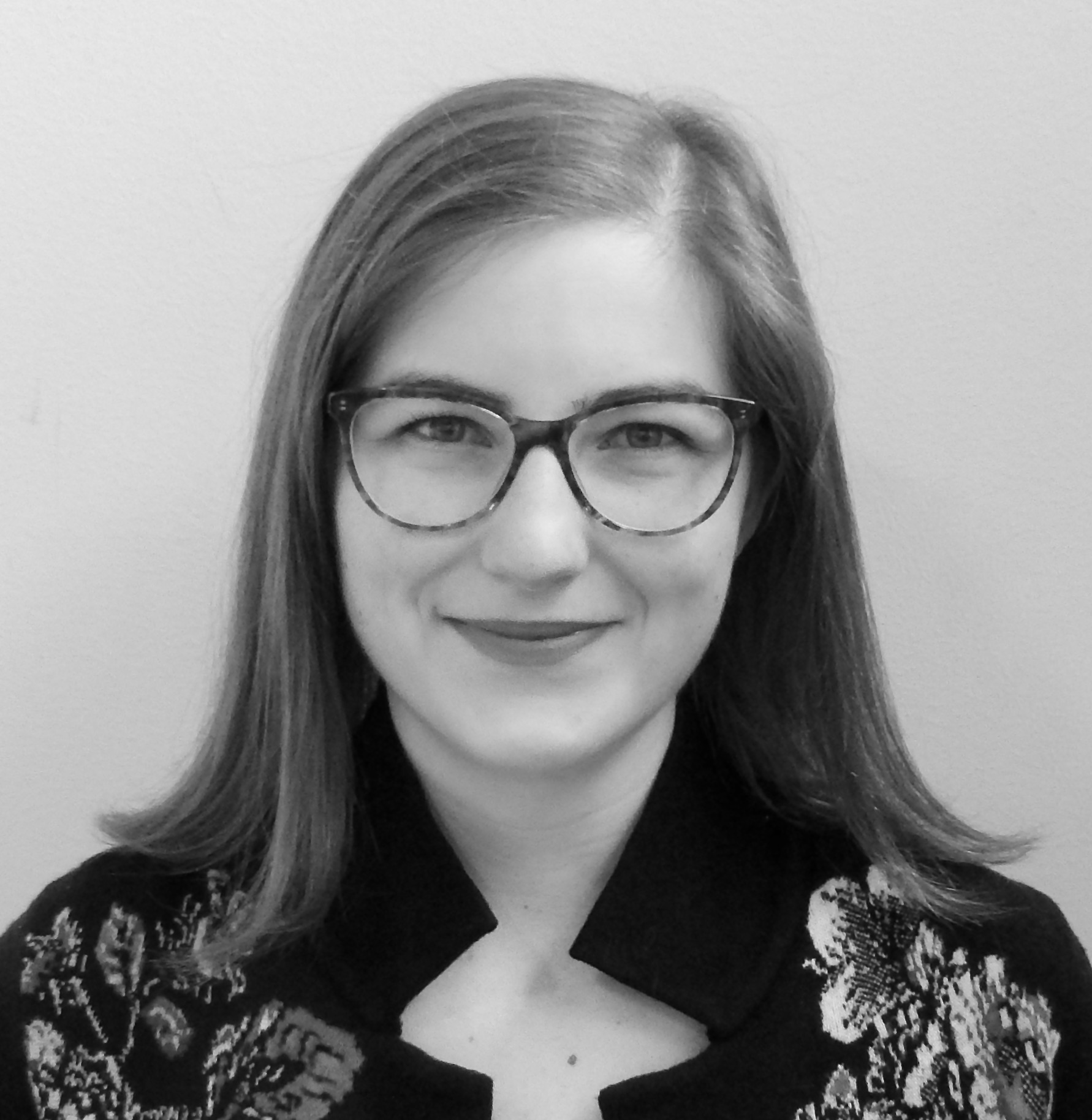}}]
{Rachel Convey} was born in the United States in 1996.  She received a M.S. degree in Speech-Language Pathology from the University of the Pacific in Stockton, California in 2019.  Her master’s thesis analyzed the impact of visual feedback on voice therapy for individuals with Parkinson’s Disease.  In the fall of 2020, she worked as a Speech-Language Pathology Clinical Fellow at Kaiser Foundation Rehabilitation Center in Vallejo, CA.  Currently, she is a Fulbright Finland grantee, conducting research at Tampere University, Finland under the direction of Nelly Penttilä, Ph.D. 
\end{IEEEbiography}
\vspace{-6mm}
\begin{IEEEbiography}[{\includegraphics[width=1in,height=1.25in,clip,keepaspectratio]{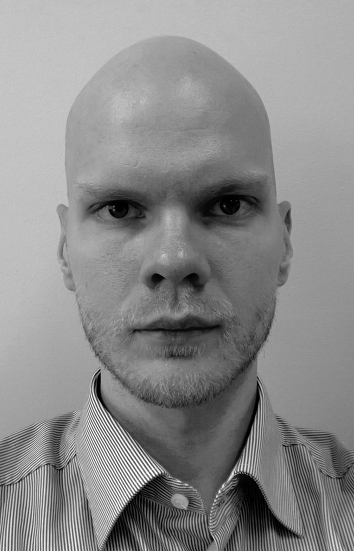}}]{Okko~R{\"a}s{\"a}nen} was born in Finland in 1984. He received the M.Sc. (Tech.) degree in language technology from the Helsinki University of Technology, Finland, in 2007, and D.Sc. (Tech.) degree in language technology from Aalto University, Finland, in 2013. He also holds the Title of Docent from Aalto University in Spoken Language Processing.

He is currently an Associate Professor at the Unit of Computing Sciences at Tampere University, Finland, and a visiting researcher at Aalto University. In 2015, he worked as a visiting researcher in the Language and Cognition Lab of Stanford University. His research interests include computational modeling of language acquisition, cognitive aspects of language processing, and speech analysis and processing in general. 
\end{IEEEbiography}

\end{document}